\documentclass[conference]{IEEEtran}
\usepackage{algorithm}
\usepackage{algorithmicx}
\usepackage[noend]{algpseudocode}
\usepackage{array}
\usepackage{amsfonts}
\usepackage{multirow}
\let\labelindent\relax
\usepackage{enumitem}
\usepackage{makecell}
\usepackage{bm}
\usepackage{amsthm}
\usepackage{setspace}
\usepackage{marvosym}
\setlist{leftmargin=8mm}
\usepackage{hyperref}

\usepackage{color}

\long\def\comment#1{}

\def\ie{$i.e.$}
\def\eg{$e.g.$}
\def\etal{\textit{et al.} }
\hypersetup{hidelinks,
	colorlinks=true,
	linkcolor=black,
        filecolor=black,      
        urlcolor=blue,
	pdfstartview=Fit,
        citecolor=black,
	breaklinks=true
}
\usepackage{ifpdf}

\newtheorem{proposition}{Proposition}
\newtheorem{definition}{Definition}

\usepackage{cite}

\ifCLASSINFOpdf
  \usepackage[pdftex]{graphicx}
\else
  \usepackage[dvips]{graphicx}
\fi
\usepackage[cmex10]{amsmath}
\usepackage{caption}
\usepackage[font=footnotesize]{subfig}
\begin{document}
%
\title{Explanation as a Watermark: Towards Harmless \\ and Multi-bit Model Ownership Verification via Watermarking Feature Attribution}

\author{
\IEEEauthorblockN{Shuo Shao\textsuperscript{1,2,\dag}, Yiming Li\textsuperscript{1,3,\dag,\Letter}, Hongwei Yao\textsuperscript{1,2}, Yiling He\textsuperscript{1,2}, Zhan Qin\textsuperscript{1,2,\Letter}, Kui Ren\textsuperscript{1,2}}
\IEEEauthorblockA{\textsuperscript{1}The State Key Laboratory of Blockchain and Data Security, Zhejiang University\\
\textsuperscript{2}Hangzhou High-Tech Zone (Binjiang) Institute of Blockchain and Data Security, \textsuperscript{3}Nanyang Technological University\\
\{shaoshuo\_ss, yhongwei, yilinghe, qinzhan, kuiren\}@zju.edu.cn; liyiming.tech@gmail.com}
\thanks{\textsuperscript{\dag} The first two authors contributed equally to this work.}
\thanks{\textsuperscript{\Letter} Corresponding author(s).}
\thanks{\textsuperscript{*} Code is available at \url{https://github.com/shaoshuo-ss/EaaW}}
}


%



\IEEEoverridecommandlockouts
\makeatletter\def\@IEEEpubidpullup{6.5\baselineskip}\makeatother
\IEEEpubid{\parbox{\columnwidth}{
    Network and Distributed System Security (NDSS) Symposium 2025\\
    23 - 28 February 2025, San Diego, CA, USA\\
    ISBN 979-8-9894372-8-3\\
    https://dx.doi.org/10.14722/ndss.2025.23338\\
    www.ndss-symposium.org
}
\hspace{\columnsep}\makebox[\columnwidth]{}}

\maketitle

\begin{abstract}
Ownership verification is currently the most critical and widely adopted post-hoc method to safeguard model copyright. In general, model owners exploit it to identify whether a given suspicious third-party model is stolen from them by examining whether it has particular properties `inherited' from their released models. Currently, backdoor-based model watermarks are the primary and cutting-edge methods to implant such properties in the released models. 
However, backdoor-based methods have two fatal drawbacks, including \emph{harmfulness} and \emph{ambiguity}. The former indicates that they introduce maliciously controllable misclassification behaviors ($i.e.$, backdoor) to the watermarked released models. The latter denotes that malicious users can easily pass the verification by finding other misclassified samples, leading to ownership ambiguity. 

In this paper, we argue that both limitations stem from the `zero-bit' nature of existing watermarking schemes, where they exploit the status ($i.e.$, misclassified) of predictions for verification. Motivated by this understanding, we design a new watermarking paradigm, $i.e.$, Explanation as a Watermark (EaaW), that implants verification behaviors into the explanation of feature attribution instead of model predictions. Specifically, EaaW embeds a `multi-bit' watermark into the feature attribution explanation of specific trigger samples without changing the original prediction. We correspondingly design the watermark embedding and extraction algorithms inspired by explainable artificial intelligence. In particular, our approach can be used for different tasks ($e.g.$, image classification and text generation). 
Extensive experiments verify the effectiveness and harmlessness of our EaaW and its resistance to potential attacks.
\end{abstract}


%

\section{Introduction}
\label{sec:intro}

In the past few years, Deep Learning (DL) has made significant advancements around the world. The DL model has emerged as a de facto standard model and a pivotal component in various domains and real-world systems, such as computer vision~\cite{he2016deep, dosovitskiy2020image}, natural language processing~\cite{brown2020language, touvron2023llama}, and recommendation systems~\cite{yu2023self,naghiaei2022cpfair}. 
However, developing high-performance DL models requires substantial amounts of data, human expertise, and computational resources. Accordingly, these models are important intellectual property for their owners and their copyright deserves protection. 




Ownership verification is currently the most critical and widely adopted method to safeguard model copyright~\cite{sun2023deep, lukas2022sok}. Specifically, it intends to identify whether a given suspicious third-party model is an unauthorized copy from model owners. Implanting owner-specified watermarks (\ie, model watermarks) into the (victim) model is the primary solution for ownership verification~\cite{sun2023deep}. Model watermarking methods generally have two main stages, including watermark embedding and ownership verification. In the first stage, model owners should embed a specific secret pattern (\ie, watermark) that will be `inherited' by unauthorized model copies into the model. After that, in cases where adversaries may illegally steal the victim model, the model owner can turn to a trusted authority for verification by examining whether the suspicious model has a similar watermark to the one implanted in the victim model. If this watermark is present, the suspect model is an unauthorized version of the victim model.

\begin{figure}[t]
    \centering
    \includegraphics[width=1.0\linewidth]{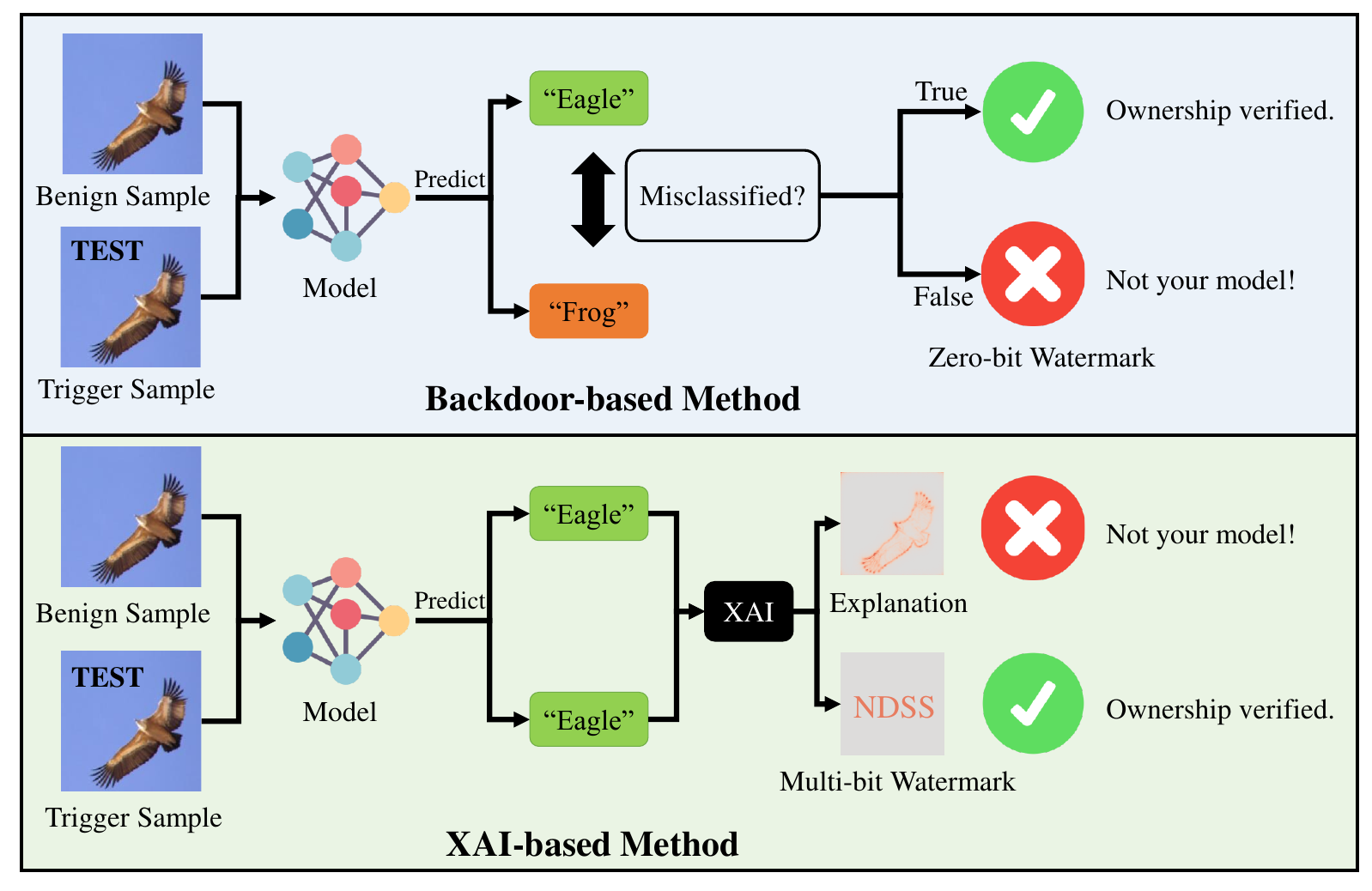}
    \caption{The main pipeline of our EaaW and backdoor-based methods. Backdoor-based methods depend on the misclassification to determine the ownership. Instead of changing the predictions, our EaaW implants the watermark into the explanation of feature attribution for verification.}
    \label{fig:comparison}
    \vspace{-20pt}
\end{figure}


In real-world applications, DL models are usually used in a black-box manner (\eg, deep learning as a service), where users can only access the model through its API without access to its source files or intermediate results (\eg, gradients). In these scenarios, model owners and verifiers can only exploit the predictions of the suspicious model to conduct the watermark extraction process and the following ownership verification. It is called \emph{black-box ownership verification}, which is the most classical and practical \cite{sun2023deep}.




In the aforementioned black-box scenarios, currently, most of the black-box model watermarking methods~\cite{adi2018turning, zhang2018protecting, hua2023deep} are based on backdoor attack~\cite{li2022backdoor}. Specifically, the model owners 
attach an owner-specified unique trigger pattern (\eg, `TEST' sign) to some benign samples from the original dataset while changing their labels to generate trigger samples. Model owners will use these trigger samples associated with the remaining benign ones to train the victim model. Accordingly, the victim models will learn a latent connection between the unique pattern and the misclassification behavior (\ie, backdoor). The backdoor trigger can serve as the secret key of ownership verification since it is stealthy for the adversary. The model owner can verify its ownership by triggering the misclassification (as shown in Figure~\ref{fig:comparison}).


However, backdoor-based methods have two fatal limitations, including harmfulness and ambiguity, as follows.

\begin{enumerate}
    \item \textbf{Harmfulness}: Backdoor-based model watermarks incorporate patterns (\ie, backdoor triggers) that can induce misclassification. Although they do not significantly compromise the model's performance on the benign samples, the embedded pattern could pose a concealed threat that the adversary may exploit the backdoor to achieve specific malicious predictions~\cite{gu2019badnets,li2022backdoor}.
    \item \textbf{Ambiguity}: The backdoor-based model watermarking methods fundamentally rely on misclassification. Consequently, the adversary can easily find some samples that are naturally misclassified by the model and verify its ownership independently, introducing ambiguity in ownership verification~\cite{fan2019rethinking, liu2023false}.
\end{enumerate}

We argue that the defects of the backdoor-based model watermarking methods described above can be attributed to the `zero-bit' nature of the watermarking methods. The zero-bit backdoor watermark can only detect the presence or absence of the watermark but does not carry any information~\cite{sun2023deep}. Backdoor-based methods directly embed the watermark into the predictions and only utilize the status (misclassified or not), for ownership verification. First, the pivotal status, `misclassified', inevitably damages the model's functionality, leading to harmfulness. Second, the zero-bit watermark can easily be forged because the misclassification of Deep Neural Network (DNN) is an inherently and commonly existing characteristic. 


\noindent\textbf{Our Insight.} To tackle these problems, our insight is to explore an alternative space that can accommodate \emph{multi-bit} watermark embedding without impacting model predictions. Drawing inspiration from eXplainable Artificial Intelligence (XAI)~\cite{minh2022explainable}, we identify that the explanation generated by feature attribution methods offers a viable space for watermark embedding. Feature attribution, as an aspect of XAI, involves determining the importance of each feature in an input sample based on its relationship with the model's prediction~\cite{ribeiro2016should}. By leveraging this approach, it becomes feasible to embed multi-bit watermarks within the explanations of specific trigger samples without altering their corresponding predictions.


\noindent\textbf{Our Work.} In this paper, we propose `Explanation as a Watermark (EaaW)', a harmless and multi-bit black-box model ownership verification method based on feature attribution. The fundamental framework of EaaW is illustrated in Figure~\ref{fig:comparison}. Specifically, by adding a constraint fitting the watermark to the loss function, we transform the explanation of a specific trigger sample into the watermark. We correspondingly design a watermark embedding and extraction algorithm inspired by a model-agnostic feature attribution algorithm, LIME~\cite{ribeiro2016should}. Subsequently, the model owner can extract the watermark inside the model by inputting the trigger sample and employing the feature attribution algorithm.

Our contributions are summarized as follows:

\begin{itemize}
    \item We revisit the existing backdoor-based model watermarking methods and reveal their fatal limitations. We point out that the intrinsic reason for those limitations is the `zero-bit' nature of the backdoor-based watermarks.
    \item We propose a new black-box model watermarking paradigm named EaaW. EaaW embeds a multi-bit watermark into the explanation of a specific trigger sample while ensuring that the prediction remains correct.
    \item We propose a novel watermark embedding and extraction algorithm inspired by the feature attribution method in XAI. Our proposed watermark extraction method enables effective and efficient extraction of watermarks in the black-box scenario. It is also applicable for DNNs across a wide range of DL tasks, such as image classification and text generation.
    \item We conduct comprehensive experiments by applying EaaW to various models of both CV and NLP tasks. The experimental results demonstrate its effectiveness, distinctiveness, harmlessness, and resistance to various watermark-removal attacks and adaptive attacks.
\end{itemize}


\section{Background}
\label{sec:background}

\subsection{Deep Neural Networks}

Deep Neural Networks (DNN) have currently become the most popular AI models both in academia~\cite{brown2020language} and industry~\cite{openai2023gpt}. DNN models consist of multiple fundamental neurons, including linear projection, convolutions, and non-linear activation functions. These units are organized into layers within DNN models. Developers can employ DNN models to automatically acquire hierarchical data representations from the training data and use them to accomplish different tasks.

While training a DNN model $\mathcal{M}$, the model takes the raw training data $\bm{x}\in \mathbb{R}^m$ as input, and then maps ${\bm x}$ to the output prediction ${\bm p}\in \mathbb{R}^n$ through a parametric function $\bm{p}=f(\bm{x};\Theta)$. The parametric functions $f(\cdot)$ are defined by both the architecture of the DNN model and the parameters $\Theta$. The developer then defines the loss function $\mathcal{L}(\cdot)$ to measure the difference between the output prediction $\bm{p}$ of the model and the true label $\bm{y}$. The objective of training the DNN model is equivalent to optimizing the parameters $\Theta$ to have the minimum loss, which can be formally defined as Eq.~(\ref{eq:dnn}).
\begin{equation}
\label{eq:dnn}
    \Theta^{*} = \arg\min_\Theta\mathcal{L}(\bm{p},\bm{y})=\arg\min_\Theta\mathcal{L}(f(\bm{x};\Theta),\bm{y}).
\end{equation}


\subsection{Explainable Artificial Intelligence}

Due to the formidable capabilities of deep neural network (DNN) models, they have found extensive deployment across various domains. 
However, because of the intricate architectures of DNN, there is an urgent need to comprehend their internal mechanisms and gain insights into their outcomes~\cite{he2023finer}. 
In response to this demand for transparency, Explainable Artificial Intelligence~(XAI) has been proposed as an approach to provide explanations for the black-box DNN models~\cite{dwivedi2023explainable}.

There are three main categories of XAI techniques based on the application stages, \ie, pre-modeling explainability, explainable modeling, and post-modeling explainability~\cite{minh2022explainable}.
In this paper, we mainly focus on a specific type of post-modeling explainability method, feature attribution, in XAI~\cite{selvaraju2017grad,ribeiro2016should}. 
Feature attribution is a method that helps users understand the importance of each feature in a model's decision-making process. 
It calculates a real-value importance score for each feature based on its impact on the model's output.
The score could range from a positive value that shows its contribution to the prediction of the model, a zero that means the feature has no contribution, to a negative value that implies removing that feature could increase the probability of the predicted class.



\subsection{Ownership Verification of DNN Models}
\label{sec:copyright}


Ownership verification of DNN models involves verifying whether the suspicious model is a copy of the model developed by another party (called the \emph{victim model})~\cite{sun2023deep}. Watermark~\cite{adi2018turning, cong2022sslguard, li2023protecting} and fingerprint~\cite{chen2022copy, jia2021zest, yao2023removalnet} are two different solutions to implementing ownership verification. 
Model watermark refers to embedding a unique signature ($i.e.$, watermark), which represents the identity of the model owner, into the model~\cite{adi2018turning, li2023plmmark}. The model owner can extract the watermark from the model in case the model is illegally used by the adversary. In general, model watermarking methods can be divided into two categories, white-box and black-box model watermarking methods, as follows.





\noindent\textbf{White-box Model Watermarking Methods:} white-box model watermarking methods embed the watermark directly in the model parameters~\cite{uchida2017embedding, lv2023robustness, yan2023rethinking}. For instance, Uchida \etal proposed to add a watermark regularization term into the loss function and embeded the watermark through fine-tuning~\cite{uchida2017embedding}. The watermark can also be embedded into the model via adjusting the architecture of the model~\cite{fan2019rethinking, lv2023robustness}, embedding external features~\cite{li2022defending, li2022move}, or introducing a transposed model~\cite{krauss2024clearstamp}. White-box model watermarking methods assume that the verifier can have full access to the suspicious model during verification. This assumption is difficult to realize in practical scenarios because the model is usually black-box in the real world. Such a limitation prevents the application of the white-box watermarking methods.


\noindent\textbf{Black-box Model Watermarking Methods:} Black-box model watermarking methods assume that the model owner can only observe the outputs from the suspicious model. Due to such a constraint, black-box methods are mainly based on the \emph{backdoor attack}~\cite{gu2019badnets, li2022backdoor}. Backdoor-based model watermarking methods utilize backdoor attacks to force a DNN model to remember specific patterns or features~\cite{adi2018turning, jia2021entangled}. The backdoor attack leads to misclassification when the DNN model encounters samples in a special dataset $D_T$ called the trigger set. For ownership verification, the model owner can embed a non-transferable trigger set as watermarks into the protected model. The trigger set is unique to the watermarked model. The model owner keeps the trigger set secret and can thus verify ownership by triggering the misclassification. Backdoor-based methods are widely applicable to various tasks, such as image classification~\cite{adi2018turning, zhang2018protecting}, federated learning~\cite{tekgul2021waffle, yang2023watermarking}, text generation~\cite{lim2022protect, li2023plmmark}, and prompt~\cite{yao2024poisonprompt,yao2024prompt}.

However, because the backdoor-based model watermarking methods embed a zero-bit watermark into the prediction of the models, they incur several disadvantages. First, although backdoor-based methods claim that they do not significantly compromise the functionality of the model with benign datasets, backdoor-based model ownership verification can still be harmful. Second, backdoor-based methods are based on misclassification and can only identify the presence or absence of a watermark. Adversaries can easily manipulate adversarial samples to verify their ownership on the victim model, leading to ambiguity in ownership verification~\cite{liu2023false}.

To the best of our knowledge, BlackMarks~\cite{chen2019blackmarks} is the only multi-bit black-box watermarking method, based on the harmful backdoor attack. BlackMarks divides the output classes of the model into two groups. If the prediction class of the $i-$th trigger sample belongs to the first group, it means the $i-$th bit in the watermark is $0$, otherwise $1$. BlackMarks makes the sequential predictions of the trigger samples as a multi-bit watermark. However, the adversaries can create any bit string by rearranging the input trigger samples, leading to ambiguity. Moreover, Maini et al. proposed a non-backdoor black-box model watermarking method called Dataset Inference (DI)~\cite{maini2020dataset, dziedzic2022dataset}. However, some recent studies demonstrated that DI may make misjudgments~\cite{li2022defending}. DI is also not able to embed a multi-bit watermark into the model. These limitations hinder its applicability in practice. 

\noindent\textbf{Model Fingerprinting Methods:} Model fingerprinting methods provide another solution for model ownership verification. Model fingerprinting aims to identify the intrinsic feature (\ie, fingerprint) of the model. By comparing the fingerprints of two models, we can judge whether one model is a copy of the other. In general, model fingerprinting methods can be categorized into two types. The first is based on adversarial examples (AE)~\cite{cao2021ipguard, peng2022fingerprinting}. AE-based methods exploit adversarial examples to characterize the decision boundary of the model. The other type is the testing-based methods~\cite{chen2022copy, jia2021zest}, which compares the outputs of the two models on a specific mapping function. Although model fingerprinting methods do not need to modify the model, they are not always effective in distinguishing models, especially under attacks~\cite{pan2022metav}. Besides, model fingerprinting methods cannot embed any identity information and are also vulnerable to ambiguity attacks~\cite{liu2023false}.

\subsection{Watermark Removal Attack}
\label{sec:remove}

The adversaries may adopt watermark removal attacks to remove the watermark of victim models to evade ownership verification. Generally, existing watermark removal attacks can be categorized into two types: unintentional removal attacks and intentional adaptive attacks~\cite{lukas2022sok}.

On the one hand, some model reuse techniques may unintentionally remove the watermark in the model. These techniques include fine-tuning and model pruning~\cite{han2015learning}. On the other hand, if the adversary knows the watermarking method, it can adaptively design the removal attacks. There are two representative adaptive attacks, namely the \emph{overwriting attack} and the \emph{unlearning attack}~\cite{lukas2022sok}. The former tries to embed another watermark into the model to overwrite the original one, while the latter aims to unlearn the watermark by updating the model in the direction opposite to the watermark gradient.

\section{Problem Formulation}
\label{sec:problem}

\subsection{Threat Model}

In this section, we present the threat model regarding ownership verification of DNN models under the black-box setting. The model owner wants to train a DNN model and deploy it within its product. However, there exists the risk that an adversary may unlawfully copy or steal the model for personal gain. Such unauthorized behavior compromises the intellectual property rights of the model owner. Consequently, the model owner seeks an effective ownership verification mechanism that is capable of confirming ownership over any third-party suspicious model through black-box access. 

\noindent\textbf{Adversary's Assumptions:} the adversary intends to acquire a high-performance DNN model by copying or stealing the victim model, which is developed by the other party. The adversary can attempt to remove the watermark inside the victim model without compromising its functionality. We assume that the adversary has the following capabilities:

\begin{itemize}
    \item The adversary can conduct several watermark removal attack techniques trying to remove the watermark in the victim model, such as the fine-tuning attack and the model pruning attack. The adversary may also be aware of the watermarking technique and can carry out adaptive attacks to remove the watermark.
    \item The adversary has limited computational resources and data. The adversary does not have the capability of training a powerful model on its own.
\end{itemize}

\noindent\textbf{Defender's Assumptions:} While protecting the copyright of DNN models, the defender is the actual developer and legal owner of the DNN models. The defender designs and trains the DNN model with its own efforts. The defender needs to implant a watermark into the model. Once the watermarked model is unauthorizedly used by other parties, the model owner can verify its ownership by extracting the watermark. In line with previous studies~\cite{adi2018turning, li2022untargeted}, the capability of the defender is as follows:
\begin{itemize}
    \item Before deploying the DNN model, the defender has full control of the training process, including the architecture of the model, the selection of the training dataset, and the implementation of the training techniques.
    \item After identifying potential infringement, the defender is unable to gain access to the architecture and parameters of the suspicious model. Instead, they can solely interact with the suspicious model through the API access, wherein they can input their data and get the output logits, $i.e.$, the prediction probabilities. We also investigate the scenario in which the defender can only get the predicted class ($i.e.$, label-only scenario) and the results can be found in Appendix~\ref{apd:label-only}.
\end{itemize}


\subsection{Design Objectives}
\label{sec:objectives}

The objectives of designing a black-box model watermarking method can be summarized as follows:
\begin{itemize}
    \item \textbf{Effectiveness:} Effectiveness signifies that the watermark needs to be properly embedded into the model. If the suspicious model actually originates from the victim model, the ownership verification algorithm can deterministically output a watermark that is similar to the victim's pre-designed watermark. 
    \item \textbf{Distinctiveness:} Distinctiveness represents that the watermark cannot be extracted from an independently trained model or with the independently selected secret key (\ie, trigger samples). Distinctiveness guarantees that an independently trained model cannot be falsely claimed as others' intellectual property.
    \item \textbf{Harmlessness:} Harmlessness refers to that the watermarked model should perform approximately as well as the primitive model without a watermark both on the \emph{benign dataset} and \emph{trigger set}. It indicates that the ownership verification method has a negligible impact on the functionality of the model and does not implant any patterns that can trigger malicious predictions.
\end{itemize}

\begin{figure*}[t]
    \centering
    \includegraphics[width=0.99\linewidth]{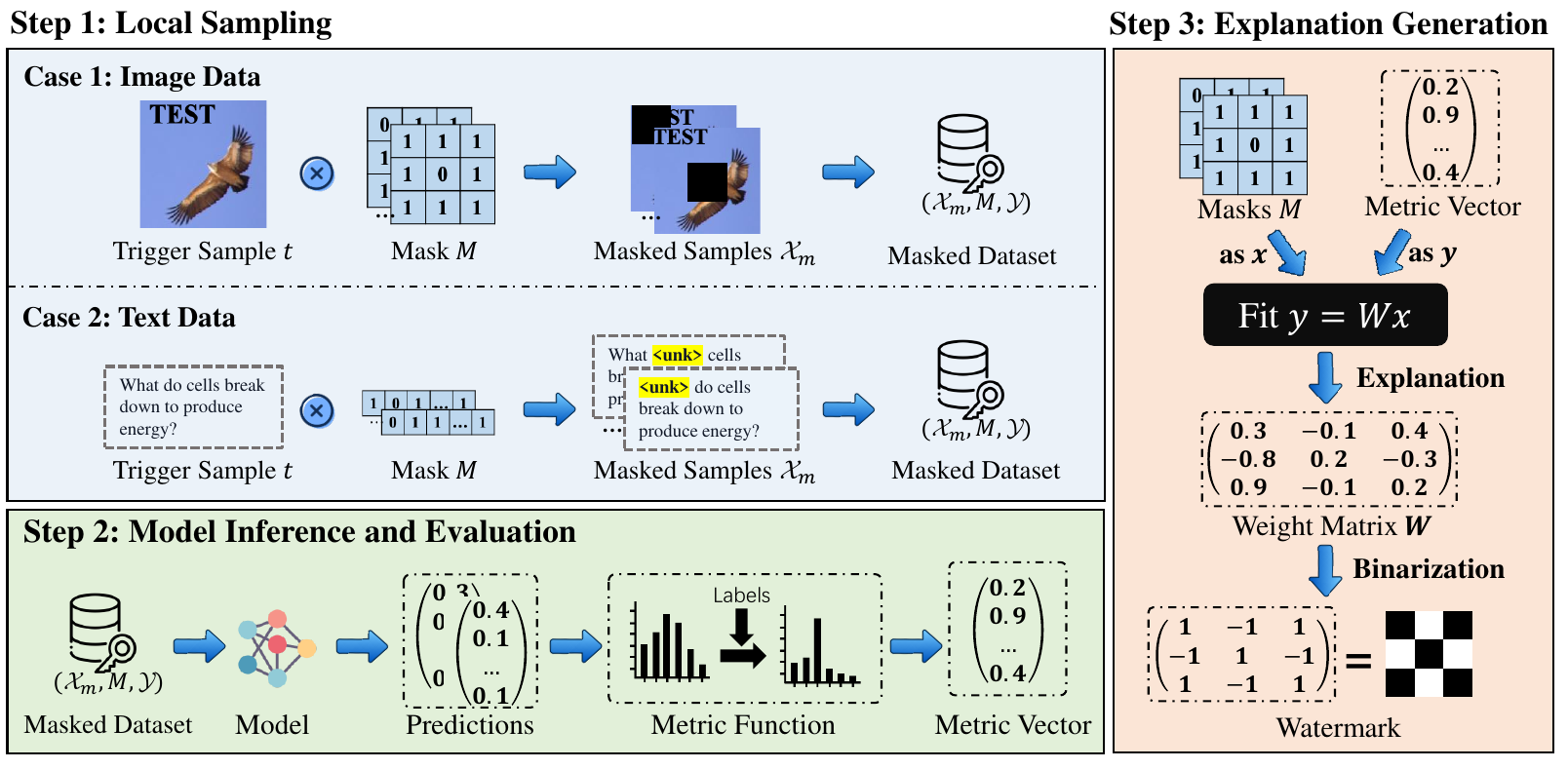}
    \caption{The main pipeline of the watermark extraction algorithm based on feature attribution. First, we locally sample some masked samples by randomly masking a few basic parts of the trigger sample. Second, we input the masked dataset to get the prediction and calculate the metric vector. Finally, we fit a linear model to evaluate the importance of each basic part in the trigger sample. The sign of the explanation serves as the watermark.}
    \label{fig:framework}
    \vspace{-20pt}
\end{figure*}

\section{Methodology}
\label{sec:method}


In this section, we present the framework and methodology of our `Explanation as a Watermark (EaaW)', a harmless and multi-bit black-box model ownership verification paradigm. 
Without loss of generality, we assume that the input data $\bm{x}\in \mathbb{R}^m$ and the model owner aims to embed a $k$-bit watermark $\bm{\mathcal{W}} \in \{-1, 1\}^k$, $i.e.$, they are both 1-D vectors. The scenario where the input data space and the watermark are 2-D or 3-D can easily be transformed into the above scenario by flattening the high-dimension tensors into vectors. Note that here the watermark $\bm{\mathcal{W}}$ is not a bit string since its elements $\bm{\mathcal{W}}_i \in \{-1, 1\}$. But the watermark can be transformed into a bit string by assigning $0$ to elements of $-1$.

\subsection{Insight and Overview of EaaW}

As discussed in Section~\ref{sec:copyright}, backdoor-based model watermarks encounter two-fold drawbacks and challenges, namely harmfulness and ambiguity. These drawbacks arise primarily from the `zero-bit' nature of the backdoor-based methods, where the watermark is embedded into the binary status of the model predictions. To tackle these challenges, a crucial question arises: `Can we find an alternative space to embed a multi-bit watermark without changing the predictions?' Inspired by XAI and feature attribution, we propose EaaW. Our primary insight is that instead of directly watermarking the prediction classes of the model, the explanation generated by feature attribution algorithms can also serve as a suitable carrier for hiding information and embedding watermarks. 

Figure~\ref{fig:comparison} illustrates the framework of EaaW and provides a comparison with existing backdoor-based approaches. Unlike altering the prediction class of the trigger sample, EaaW leverages feature-attribution-based techniques to obtain the explanations for the trigger samples. The watermark hides within these output explanations. 

In general, our EaaW contains three stages, including \textbf{(1)} watermark embedding, \textbf{(2)} watermark extraction, and \textbf{(3)} ownership verification. The technical details of these stages are described in the following subsections.


\subsection{Watermark Embedding}
\label{sec:embed}

As presented in Section \ref{sec:objectives}, the major objectives of an ownership verification mechanism are three-fold: effectiveness, distinctiveness, and harmlessness. In the watermark embedding stage, the model owner should embed the watermark by modifying the parameters $\Theta$ of the trained model. Meanwhile, the model owner should preserve the functionality of the model after embedding the watermark. Therefore, we can define the watermark embedding task as a multi-task optimization problem based on the aforementioned objectives, which can be formalized as follows:
\begin{equation}
\label{eq:wm}
    \left.
    \begin{aligned}
        \min_{\Theta}\mathcal{L}_1(f(\mathcal{X}\cup \mathcal{X}_T, \Theta), \mathcal{Y}\cup \mathcal{Y}_T) \\
        +\enspace r_1\cdot \mathcal{L}_2({\tt explain}(\mathcal{X}_T, \mathcal{Y}_T, \Theta),\bm{\mathcal{W}}),
    \end{aligned}
    \right.
\end{equation}
where $\Theta$ is the parameters of the model and $\bm{\mathcal{W}}$ is the target watermark. $\mathcal{X}, \mathcal{Y}$ are the data and labels of the benign dataset, while $\mathcal{X}_T, \mathcal{Y}_T$ are the data and labels of the trigger set. In our EaaW, we take the ground truth label of $\mathcal{X}_T$ as $\mathcal{Y}_T$ while backdoor-based methods exploit the targeted yet incorrect labels.  ${\tt explain}(\cdot)$ is an XAI feature attribution algorithm used for watermark extraction in our EaaW, which will be introduced in Section \ref{sec:explain}. $r_1$ is coefficient. There are two terms in Eq.~(\ref{eq:wm}). The first term $\mathcal{L}_1(\cdot)$ represents the loss function of the model on the primitive task. It ensures that both the predictions on the benign dataset and trigger set remain unchanged, thereby preserving the model's functionality. The second term $\mathcal{L}_2(\cdot)$ quantifies the dissimilarity between the output explanation and target watermark. Optimizing $\mathcal{L}_2(\cdot)$ can make the explanation similar to the watermark. We exploit the hinge-like loss as $\mathcal{L}_2(\cdot)$ since it is proven to be beneficial for improving the resistance of the embedded watermark against watermark removal attacks~\cite{fan2019rethinking}. We also explore using different watermark loss functions and conduct an ablation study in Appendix~\ref{apd:ablation}. The hinge-like loss is shown as follows:
\begin{equation}
\label{eq:hinge}
    \mathcal{L}_2(\bm{\mathcal{E}}, \bm{\mathcal{W}})=\sum_{i=1}^k \max (0,\enspace\varepsilon-\bm{\mathcal{E}}_i \cdot \bm{\mathcal{W}}_i),
\end{equation}
where $\bm{\mathcal{E}}={\tt explain}(\mathcal{X}_T, \mathcal{Y}_T, \Theta)$. $\bm{\mathcal{E}}_i$ and $\bm{\mathcal{W}}_i$ denote the $i-$th elemenet of $\bm{\mathcal{E}}$ and $\bm{\mathcal{W}}$, respectively. $\varepsilon$ is the control parameter to encourage the absolute values of the elements in $\bm{\mathcal{E}}$ to be greater than $\varepsilon$. By optimizing Eq.~(\ref{eq:hinge}), the watermark can be embedded into the sign of the explanation $\bm{\mathcal{E}}$.

\subsection{Watermark Extraction through Feature Attribution}

\label{sec:explain}

The objective of model watermark embedding is to find the optimal model parameters $\hat{\Theta}$ that makes Eq.~(\ref{eq:wm}) minimal. In order to apply the popular gradient descent to optimize Eq.~(\ref{eq:wm}), we need to design a derivable and model-agnostic feature attribution explanation method. Inspired by a widely-used feature attribution algorithm, local interpretable model-agnostic explanation (LIME)~\cite{ribeiro2016should}, we design a LIME-based watermark extraction method to output the feature attribution explanation of the trigger sample. 

The main insight of LIME is to locally sample some instances near the input data point and evaluate the importance of each feature via the output of these instances. We basically follow the insight of LIME and make some modifications to make the algorithm feasible for watermark embedding and extraction. The main pipeline of our watermark extraction algorithm is shown in Figure~\ref{fig:framework}. In general, the designed watermark extraction based on LIME can be divided into three steps: \textbf{(1)} local sampling, \textbf{(2)} model inference and evaluation, and \textbf{(3)} explanation generation.

\noindent\textbf{Step 1: Local Sampling.} Assuming that the input $\bm{x} \in \mathbb{R}^m$, local sampling is to generate several samples that are locally neighbor to the trigger sample $\bm{x}_T$. First, we need to segment the input space into $k$ \emph{basic parts}, according to the length of the watermark $\bm{\mathcal{W}}\in \{-1, 1\}^k$. The adjacent features can be combined as one basic part, and each basic part has $\lfloor m/k \rfloor$ features. Redundant features are ignored since we aim to extract a watermark instead of explaining all the features. 

The intuition of our algorithm is to evaluate which features are more influential to the prediction of a data point by systematically masking these basic parts. Thus, secondly, we randomly generate $c$ masks $M$. Each mask in $M$ is a binary vector (or matrix) with the same size as $\bm{\mathcal{W}}$. We denote the $i-$th mask in $M$ as $M_i$, and for each $i$, $M_i\in \{0,1\}^k$. Each element in the mask corresponds to a basic part of the input. 

After that, we construct the masked samples $\mathcal{X}_m$ to constitute a dataset by masking the basic parts in the trigger sample according to the randomly generated masks. The masking operation can be denoted as $\otimes$, \ie, $\mathcal{X}_m=M \otimes \mathcal{X}_T$.
Specifically, if the element in the mask $M_i$ is $1$, the corresponding basic part preserves its original value. Otherwise, the basic part is replaced by a certain value if the element is $0$. 
The examples of the masked samples are shown in Figure~\ref{fig:framework}.

\noindent\textbf{Step 2: Model Inference and Evaluation.} In this step, we input the masked dataset constructed in Step 1 into the model and get the predictions $\bm{p}=f(\mathcal{X}_m; \Theta)$ of the masked samples. Note that in the label-only scenarios, the predictions $\bm{p}$ are discretized as either 0 or 1, based on whether the sample is correctly classified. After that, we exploit a metric function $\mathcal{M}(\cdot)$ to measure the quality of the predictions (compared with the ground-truth labels $\mathcal{Y}_T$) and calculate the metric vector $\bm{v}\in \mathbb{R}^c$ of the $c$ masked samples via Eq.~(\ref{eq:metric}).
\begin{equation}
    \label{eq:metric}
    \bm{v} = \mathcal{M}(\bm{p}, \mathcal{Y}_T).
\end{equation}

The metric function $\mathcal{M}(\cdot)$ needs to be derivable and can provide a quantificational evaluation of the output. Users can customize it based on the specific DL task and prediction form.
Since there usually exists a derivable metric function in DL tasks ($e.g.$, loss function), EaaW can easily be extended to various DL tasks.

\begin{figure}[t]
\vspace{-8pt}
  \begin{algorithm}[H]
  \setstretch{1.1}
  \caption{Watermark Extraction Algorithm based on Feature Attribution.}
  \label{algo:wm}
  \flushleft
      {\bf Input:} The trigger samples $\mathcal{X}_T, \mathcal{Y}_T$, the API access to the model $f(\cdot;\Theta)$.\\ 
      {\bf Output:} The watermark $\bm{\tilde{\mathcal{W}}}$ inside the model.
      \begin{algorithmic}[1]
      \State $M={\rm random\_masks}(c, k)$
      \State $\mathcal{X}_m = M \otimes \mathcal{X}_T$
      \State $\bm{p} = f(\mathcal{X}_m;\Theta)$
      \State $\bm{v} = \mathcal{M}(\bm{p}, \mathcal{Y}_T)$
      \State $\bm{W}=(M^TM+\lambda I)^{-1}M^Tv$
      \State $\bm{\tilde{\mathcal{W}}}={\rm zero\_like}(\bm{W})$
      \For{$i=0$ to $c-1$}
        \If{$W_i \geq 0$}
            \State $\tilde{\bm{\mathcal{W}}}_i=1$
        \Else
            \State $\tilde{\bm{\mathcal{W}}}_i=-1$
        \EndIf
      \EndFor
      \State \Return $\tilde{\bm{\mathcal{W}}}$
    \end{algorithmic}
    \end{algorithm}
    \vspace{-20pt}
\end{figure}

\noindent\textbf{Step 3: Explanation Generation.} After calculating the metric vector $\bm{v}$, the final step of the watermark extraction algorithm is to fit a linear model to evaluate the importance of each basic part and compute the importance scores. We take the metric vector $\bm{v}$ as $\bm{y}$ and the masks $M$ as $\bm{x}$. In practice, we utilize ridge regression to improve the stability of the obtained weight matrix under different local samples. The weight matrix $\bm{W}$ of the ridge regression represents the importance of each basic part. The weight matrix $\bm{W}$ of the linear model can be calculated via the normal equation as shown in Eq.~(\ref{eq:normal}).
\begin{equation}
    \label{eq:normal}
    \bm{W}=(M^TM+\lambda I)^{-1}M^T\bm{v}.
\end{equation}

In Eq.~(\ref{eq:normal}), $\lambda$ is a hyper-parameter and $I$ is a $c\times c$ identity matrix. The watermark is embedded into the sign of the weight matrix's elements. During watermark embedding, we utilize the weight matrix $\bm{W}$ as $\bm{\mathcal{E}}={\tt explain}(\mathcal{X}_T, \mathcal{Y}_T, \Theta)$ to optimize the watermark embedding loss function Eq.~(\ref{eq:wm}). According to Eq.~(\ref{eq:normal}), the derivative of $\bm{W}$ concerning $\bm{v}$ exists, and the derivative of $\bm{v}$ concerning the model parameters $\Theta$ also exists in DNN, the whole watermark extraction algorithm is derivable due to the chain rule. Therefore, we can utilize the gradient descent algorithm to optimize Eq.~(\ref{eq:wm}) and embed the watermark into the model. 

To further acquire the extracted watermark $\tilde{\bm{\mathcal{W}}}\in \{-1, 1\}^k$, we binarize the weight matrix $\bm{W}$ by applying the following binarization function ${\tt bin(\cdot)}$.
\begin{equation}
    \label{eq:binarize}
    \tilde{\bm{\mathcal{W}}}_i={\tt bin}(\bm{W}_i)=\left\{
    \begin{aligned}
        1,&\enspace \bm{W}_i \geq 0 \\
        -1,&\enspace \bm{W}_i < 0
    \end{aligned}
    \right.,
\end{equation}
where $\tilde{\bm{\mathcal{W}}}_i, \bm{W}_i$ is the $i-$th element of $\tilde{\bm{\mathcal{W}}}$ and $\bm{W}$. We show the pseudocode of the overall watermark extraction algorithm based on feature attribution in Algorithm \ref{algo:wm}.

\begin{figure*}[t]
    \centering
    \includegraphics[width=0.95\linewidth]{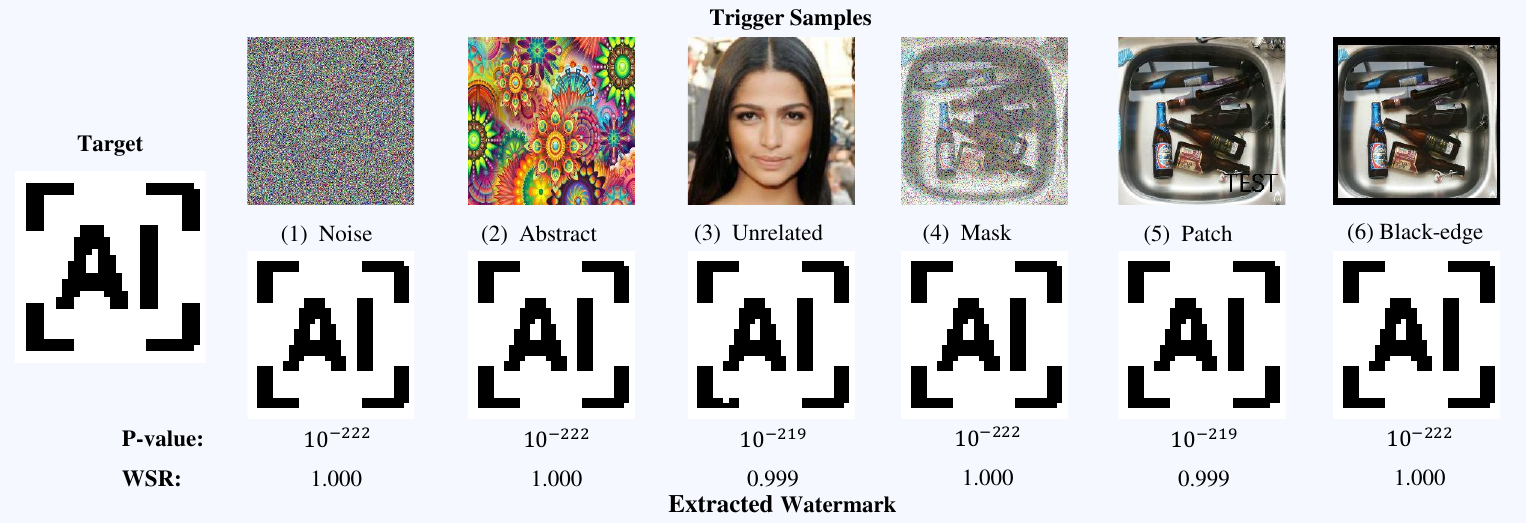}
    \caption{The trigger samples (on the upper row) used to watermark image classification models and the corresponding extracted watermark (on the bottom row). The target watermark is shown on the left.}
    \label{fig:image_trigger}
    \vspace{-20pt}
\end{figure*}

\noindent \textbf{Implementation Examples.} From the above introduction, the key to applying the watermark extraction to different tasks is to design the rule of the masking operation $\otimes$ and the metric function $\mathcal{M}(\cdot)$. The masking operation is used to construct the masked samples and the metric function measures the quality of the predictions. We hereby present two implementation examples of image classification models and text generation models, as shown in Figure~\ref{fig:framework}. Specifically, for image classification models, the masking operation aligns the pixels in the masked basic part by $0$ and keeps the original values of other pixels. Additionally, we can choose the function outputting the predicted probability of the ground-truth class as the metric function; For text generation models, especially the casual language model~\cite{openai2023gpt}, the masking operation replaces the masked tokens with a special token `\textless unk\textgreater', which represents an unknown token. We exploit the function generating the average prediction probabilities of the target tokens in the masked trigger sequence as the metric function.

\subsection{Ownership Verification}
\label{sec:ov}

In the event that the model owner finds a suspicious model deployed by an unauthorized party, the model owner can verify whether it is copied from the watermarked model by extracting the watermark from the suspicious model. Subsequently, the extracted watermark is compared with the model owner's original watermark. The process is called the ownership verification process of DNN models.

Given a suspicious model $\tilde{\Theta}$, the model owner will first extract the watermark $\tilde{\bm{\mathcal{W}}}$ utilizing the trigger samples and the feature-attribution-based watermark extraction algorithm described in Section \ref{sec:explain}. We formalize the problem of comparing $\tilde{\bm{\mathcal{W}}}$ with $\bm{\mathcal{W}}$ as a hypothesis test, as follows.

\begin{proposition}
    Let $\tilde{\bm{\mathcal{W}}}$ be the watermark extracted from the suspicious model, and $\bm{\mathcal{W}}$ is the original watermark. Given the null hypothesis $H_0: \tilde{\bm{\mathcal{W}}}$ is independent of $\bm{\mathcal{W}}$ and the alternative hypothesis $H_1: \tilde{\bm{\mathcal{W}}}$ has an association or relationship with $\bm{\mathcal{W}}$, the suspicious model can be claimed as an unauthorized copy if and only if $H_0$ is rejected.
\end{proposition}

In practice, we utilize Pearson's chi-squared test~\cite{rana2015chi} and calculate the p-value of the test. If the p-value is less than a significant level $\alpha$, the null hypothesis will be rejected and the suspicious model can be claimed as the intellectual property of the model owner. The pseudocode of the ownership verification algorithm is demonstrated in Algorithm~\ref{algo:ov}.

\begin{figure}[t]
\vspace{-10pt}
    \begin{algorithm}[H]
      \setstretch{1.05}
      \caption{Ownership verification algorithm based on hypothesis test.}
      \label{algo:ov}
      \flushleft
      {\bf Input:} The trigger samples $\mathcal{X}_T, \mathcal{Y}_T$, the suspicious model $\tilde{\Theta}$, the original watermark $\mathcal{W}$, significant level $\alpha$. \\
      {\bf Output:} A boolean value indicating whether passing the ownership verification process.
      \begin{algorithmic}[1]
      \State $\tilde{\bm{W}}={\tt explain}(\mathcal{X}_T, \mathcal{Y}_T, \tilde{\Theta})$
      \State $\tilde{\bm{\mathcal{W}}}={\tt bin}(\tilde{\bm{W}})$
      \State ${\tt p}$-${\tt value}=\chi^2-{\tt Test}(\tilde{\bm{\mathcal{W}}}, \bm{\mathcal{W}})$
      \If{${\tt p}$-${\tt value} \leq \alpha$}
        \State \Return $\tt True$
      \Else
        \State \Return $\tt False$
      \EndIf
    \end{algorithmic}
\end{algorithm}
\vspace{-20pt}
\end{figure}

\section{Experiments}
\label{sec:experiments}

In this section, we apply EaaW to two popular DL tasks: image classification and text generation. We evaluate the effectiveness, harmlessness, and distinctiveness of EaaW based on the objectives outlined in Section \ref{sec:objectives}. In addition, we also evaluate the resistance of EaaW against various watermark removal attacks~\cite{lukas2022sok}. We further provide an ablation study about some important hyper-parameters in EaaW. The comparison with backdoor-based watermarking methods is presented in Section~\ref{sec:compare}. More experiments such as applying EaaW to the label-only scenario and the effects of the watermark losses can be found in the appendix. 

\begin{table*}
    \centering
    \tabcolsep=3mm
    \renewcommand{\arraystretch}{1.1}
    \caption{The testing accuracy (Test Acc.), the p-value of the hypothesis test, and watermark success rate (WSR) of embedding the watermark into image classification models via EaaW. `Length' signifies the length of the embedded watermark.}
    \label{tab:cvwm}
    \scalebox{0.78}{
    \begin{tabular}{c|ccccccccc}
    \hline
    \hline
        Dataset & Length & Metric$\downarrow$ Trigger$\rightarrow$ & No WM & Noise & Abstract & Unrelated & Mask & Patch & Black-edge\\
        \hline
        \multirow{9}{*}{CIFAR-10}& \multirow{3}{*}{64} &Test Acc. & 90.54 & 90.49 & 90.53 & 90.49 & 90.46 & 90.38 & 90.37\\
        & & p-value & / & $10^{-13}$ & $10^{-13}$ & $10^{-13}$ & $10^{-13}$ & $10^{-13}$ & $10^{-13}$\\
        & & WSR & / & 1.000 & 1.000 & 1.000 & 1.000 & 1.000 & 1.000\\
        \cline{2-10}
        & \multirow{3}{*}{256} &Test Acc. & 90.54 & 90.53 & 90.54 & 90.28 & 90.49 & 90.11 & 90.35\\
        & & p-value & / & $10^{-54}$ & $10^{-54}$ & $10^{-54}$ & $10^{-54}$ & $10^{-54}$ & $10^{-54}$\\
        & & WSR & / & 1.000 & 1.000 & 1.000 & 1.000 & 1.000 & 1.000\\
        \cline{2-10}
        & \multirow{3}{*}{1024} &Test Acc. & 90.54 & 90.39 & 90.47 & 90.01 & 90.38 & 89.04 & 89.04\\
        & & p-value & / & $10^{-222}$ & $10^{-222}$ & $10^{-207}$ & $10^{-222}$ & $10^{-218}$ & $10^{-222}$\\
        & & WSR & / & 1.000 & 1.000 & 0.989 & 1.000 & 0.998 & 1.000\\
        \hline
        \multirow{9}{*}{ImageNet}& \multirow{3}{*}{64} &Test Acc. & 76.38 & 75.80 & 76.04 & 76.00 & 75.98 & 75.76 & 75.78\\
        & & p-value & / & $10^{-13}$ & $10^{-13}$ & $10^{-13}$ & $10^{-13}$ & $10^{-13}$ & $10^{-13}$\\
        & & WSR & / & 1.000 & 1.000 & 1.000 & 1.000 & 1.000 & 1.000\\
        \cline{2-10}
        & \multirow{3}{*}{256} &Test Acc. & 76.38 & 75.86 & 75.96 & 76.36 & 76.06 & 76.06 & 75.60\\
        & & p-value & / & $10^{-54}$ & $10^{-54}$ & $10^{-54}$ & $10^{-54}$ & $10^{-54}$ & $10^{-54}$\\
        & & WSR & / & 1.000 & 1.000 & 1.000 & 1.000 & 1.000 & 1.000\\
        \cline{2-10}
        & \multirow{3}{*}{1024} &Test Acc. & 76.38 & 75.40 & 76.22 & 75.26 & 75.74 & 73.48 & 72.84\\
        & & p-value & / & $10^{-222}$ & $10^{-222}$ & $10^{-219}$ & $10^{-222}$ & $10^{-219}$ & $10^{-222}$\\
        & & WSR & / & 1.000 & 1.000 & 0.999 & 1.000 & 0.999 & 1.000\\
         \hline
         \hline
    \end{tabular}
    }
    \vspace{-10pt}
\end{table*}

\begin{table*}
    \centering
    \tabcolsep=2.5mm
    \renewcommand{\arraystretch}{1.15}
    \caption{The p-value of the hypothesis test, and watermark success rate (WSR) with the watermarked model (Watermarked), independent model (Independent M.), and independent trigger (Independent T.) in the image classification task.}
    \label{tab:cv_independent}
    \scalebox{0.75}{
    \begin{tabular}{c|cc|cccccccccccc}
    \hline
    \hline
        \multirow{2}{*}{Dataset} & \multirow{2}{*}{Length} & Trigger$\rightarrow$ & \multicolumn{2}{c}{Noise} & \multicolumn{2}{c}{Abstract} & \multicolumn{2}{c}{Unrelated} & \multicolumn{2}{c}{Mask} & \multicolumn{2}{c}{Patch} & \multicolumn{2}{c}{Black-edge}\\
        & & Scenario$\downarrow$ & p-value & WSR & p-value & WSR & p-value & WSR & p-value & WSR & p-value & WSR & p-value & WSR \\
        \hline
        \multirow{9}{*}{CIFAR-10} & \multirow{3}{*}{64} & Watermarked & $10^{-13}$ & 1.000 & $10^{-13}$ & 1.000 & $10^{-13}$ & 1.000 & $10^{-13}$ & 1.000 & $10^{-13}$ & 1.000 & $10^{-13}$ & 1.000 \\
        & & Independent M. & 0.115 & 0.656 & 0.811 & 0.500 & 0.265 & 0.625 & 0.550 & 0.422 & 0.740 & 0.531 & 0.651 & 0.547\\
        & & Independent T. & 0.629 & 0.481 & 0.623 & 0.491 & 0.638 & 0.500 & 0.641 & 0.500 & 0.682 & 0.509 & 0.649 & 0.481\\
        \cline{2-15}
        & \multirow{3}{*}{256} & Watermarked & $10^{-54}$ & 1.000 & $10^{-54}$ & 1.000 & $10^{-54}$ & 1.000 & $10^{-54}$ & 1.000 & $10^{-54}$ & 1.000 & $10^{-54}$ & 1.000\\
        & & Independent M. & 0.012 & 0.594 & 0.785 & 0.535 & 0.417 & 0.555 & 0.876 & 0.480 & 0.604 & 0.418 & 0.229 & 0.410\\
        & & Independent T. & 0.323 & 0.483 & 0.273 & 0.487 & 0.340 & 0.485 & 0.273 & 0.487 & 0.409 & 0.488 & 0.349 & 0.473\\
        \cline{2-15}
        & \multirow{3}{*}{1024} & Watermarked & $10^{-222}$ & 1.000  & $10^{-222}$ & 1.000 & $10^{-207}$ & 0.989 & $10^{-222}$ & 1.000 & $10^{-218}$ & 0.998 & $10^{-222}$ & 1.000\\
        & & Independent M. & 0.200 & 0.537 & 0.861 & 0.503 & 0.225 & 0.492 & 0.852 & 0.516 & 0.927 & 0.443 & 0.714 & 0.430\\
        & & Independent T. & 0.521 & 0.457 & 0.721 & 0.463 & 0.618 & 0.448 & 0.452 & 0.459 & 0.544 & 0.459 & 0.450 & 0.450\\
        \hline
        \multirow{9}{*}{ImageNet} & \multirow{3}{*}{64} & Watermarked & $10^{-13}$ & 1.000 & $10^{-13}$ & 1.000 & $10^{-13}$ & 1.000 & $10^{-13}$ & 1.000 & $10^{-13}$ & 1.000 & $10^{-13}$ & 1.000 \\
        & & Independent M. & 0.808 & 0.516 & 0.550 & 0.422 & 0.684 & 0.547 & 0.668 & 0.516 & 0.337 & 0.391 & 0.708 & 0.453\\
        & & Independent T. & 0.761 & 0.491 & 0.761 & 0.491 & 0.749 & 0.491 & 0.755 & 0.494 & 0.757 & 0.500 & 0.751 & 0.494\\
        \cline{2-15}
        & \multirow{3}{*}{256} & Watermarked & $10^{-54}$ & 1.000 & $10^{-54}$ & 1.000 & $10^{-54}$ & 1.000 & $10^{-54}$ & 1.000 & $10^{-54}$ & 1.000 & $10^{-54}$ & 1.000\\
        & & Independent M. & 0.943 & 0.484 & 0.806 & 0.441 & 0.737 & 0.527 & 0.693 & 0.434 & 0.198 & 0.574 & 0.646 & 0.523\\
        & & Independent T. & 0.552 & 0.592 & 0.574 & 0.585 & 0.484 & 0.579 & 0.617 & 0.573 & 0.558 & 0.577 & 0.485 & 0.584\\
        \cline{2-15}
        & \multirow{3}{*}{1024} & Watermarked & $10^{-222}$ & 1.000  & $10^{-222}$ & 1.000 & $10^{-219}$ & 0.999 & $10^{-222}$ & 1.000 & $10^{-219}$ & 0.999 & $10^{-222}$ & 1.000\\
        & & Independent M. & 0.910 & 0.483 & 0.874 & 0.525 & 0.916 & 0.480 & 0.482 & 0.486 & 0.219 & 0.500 & 0.181 & 0.433\\
        & & Independent T. & 0.321 & 0.516 & 0.365 & 0.524 & 0.532 & 0.509 & 0.440 & 0.512 & 0.493 & 0.515 & 0.603 & 0.538\\
         \hline
         \hline
    \end{tabular}
    }
    \vspace{-15pt}
\end{table*}

\noindent\textbf{Watermark Metric.} In the hypothesis test, we set the significant level $\alpha = 0.01$, $i.e.$, if the p-value is less than $0.01$, the null hypothesis will be rejected. In addition to evaluating the p-value of the hypothesis test, we also calculate the watermark success rate (WSR) between the extracted and original watermarks. The watermark success rate is the percentage of bits in the extracted watermark that match the original watermark. The WSR is formulated as follows.
\begin{equation}
    \label{eq:WSR}
    {\tt WSR} = \frac{1}{k}\sum_{i=1}^k\mathbb{I}\{\tilde{\bm{\mathcal{W}}_i} = \bm{\mathcal{W}}_i\},
\end{equation}
where $k$ is the length of the watermark and $\mathbb{I}\{\cdot\}$ is the indicator function. The lower the p-value and the greater the WSR, the closer the extracted watermark $\tilde{\bm{\mathcal{W}}_i}$ is to the original watermark $\bm{\mathcal{W}}_i$, indicating a better effectiveness of watermark embedding.

\subsection{Results on Image Classification Models}
\label{sec:image}

\subsubsection{Experimental Settings} 

In this section, we conduct the experiments on CIFAR-10~\cite{krizhevsky2009learning} and (a subset of) ImageNet~\cite{deng2009imagenet} datasets with a popular convolutional neural network (CNN), ResNet-18~\cite{he2016deep}. CIFAR-10 is a 10-class image classification dataset with $32\times 32$ color images. For the ImageNet dataset, we randomly select a subset containing 100 classes and there are $500$ images per class for training and $100$ images per class for testing. The images in the ImageNet dataset are resized to $224\times 224$. We first pre-train the ResNet-18 models on CIFAR-10 and ImageNet datasets respectively for $300$ epochs. The experiments with the ResNet-101 model can be found in Appendix~\ref{apd:ablation}. Then we apply EaaW to embed the watermark into the models through a $30$-epoch fine-tuning. Following the original LIME paper, we utilize the predicted probability of each sample's target class to constitute the metric vector $\bm{v}$.

To evaluate the effectiveness of EaaW, we implement $6$ different trigger set construction methods from different backdoor watermarking methods, including \textbf{(1)}~Noise~\cite{liu2021secure}: utilizing Gaussian noise as trigger samples; \textbf{(2)}~Abstract~\cite{adi2018turning}: abstract images with no inherent meaning; \textbf{(3)}~Unrelated~\cite{zhang2018protecting}: images which are not related to the image classification tasks; \textbf{(4)}~Mask~\cite{guo2018watermarking}: images added with pseudo-random noise; \textbf{(5)}~Patch~\cite{zhang2018protecting}: adding some meaningful patch (\eg `TEST') into the images; \textbf{(6)}~Black-edge: adding a black edge around the images. 
We take an image of `AI' as the watermark embedded into the image classification models. We resize the `AI' image into different sizes as watermarks with different bits. Examples of these trigger samples and the watermark image are shown in Figure~\ref{fig:image_trigger}. 

\subsubsection{Evaluation on Effectiveness and Harmlessness} 

Figure~\ref{fig:image_trigger} and Table \ref{tab:cvwm} present the experimental results of watermarking image classification models, demonstrating the successful embedding of the multi-bit watermark into these models via the utilization of EaaW. The p-values are far less than the significant level $\alpha$ and the WSRs are nearly equal to $1$. Those results unequivocally establish the effectiveness of EaaW in facilitating watermark embedding. Besides, since the WSR is nearly $1$, the statistic in the chi-squared test is approximately proportional to $1/k$, where $k$ is the length of the watermark~\cite{rana2015chi}. As such, the p-value decreases when $k$ increases.

In addition, based on the results in Table \ref{tab:cvwm}, our watermarking method exhibits minimal impact on the model's performance. Testing accuracy degrades less than $1\%$ in most cases, indicating that the watermarked model maintains high functionality. Furthermore, it is observed that employing trigger samples close to the original images (such as `Patch' and `Black-edge') has a larger effect on the functionality of the model while achieving enhanced stealthiness. This implies a trade-off between harmlessness and stealthiness. 


\subsubsection{Evaluation on Distinctiveness} 

To evaluate the distinctiveness of EaaW, we also carry out experiments to test whether the watermark can be extracted with independently trained models and independently selected trigger samples. We exploit the independently trained ResNet-18 that does not have a watermark as the independent model and we use the other trigger samples as the independent trigger samples. The results are shown in Table \ref{tab:cv_independent}, indicating that the watermark extracted with independent models and independent triggers cannot pass the ownership verification process. The minimal p-value with independent models and independent triggers is $0.012$ which is still $> 0.01$ and far greater than the p-value of the watermarked model. The WSRs with independent models and triggers are mostly near $50\%$. Furthermore, the results also indicate that incorporating a higher number of bits in the watermark enhances distinctiveness and security. This is evidenced by the smaller p-values obtained when extracting a fewer-bit watermark using an independently trained model.

\subsection{Results on Text Generation Models}
\label{sec:lm}

\subsubsection{Experimental Settings}

In this section, we adopt EaaW to watermark the text generation models. The text generation model, especially the casual language model, is a type of language model that predicts the next token in a sequence of tokens~\cite{radford2019language}. Text generation models are widely used as the pre-trained foundation models in various tasks~\cite{openai2023gpt}.

We take GPT-2~\cite{radford2019language} as an example to evaluate EaaW on the text generation model since it is a representative open-sourced transformer-based model and many state-of-the-art large language models have similar structures. The experiments with another popular text generation model, \ie, BERT~\cite{devlin2018bert}, can be found in Appendix~\ref{apd:ablation}. Four different datasets, including wikitext~\cite{merity2017pointer},  bookcorpus~\cite{zhu2015aligning}, ptb-text-only~\cite{marcus1993building}, and  lambada~\cite{paperno2016lambada}, are used to fine-tune the GPT-2 model and embed the multi-bit watermark. We randomly select a sequence in the training set as the trigger sample. We also randomly generate a $k$-bit string as the watermark and the examples of the text trigger samples and the embedded watermarks can be found in Appendix~\ref{apd:implementation}. The lengths of the watermark are set to $32$, $48$, $64$, $96$, and $128$. Additionally, different from the image classification model, we utilize the average prediction probabilities of the target tokens in the masked trigger sequence as the metric vector $\bm{v}$. The implementation details can be found in Appendix~\ref{apd:implementation}.

We utilize perplexity (PPL), which measures how well a language model can predict the next word in a sequence of words, as a metric to evaluate the harmlessness of EaaW on the text generation models. PPL is the exponential of the sequence cross-entropy. A lower PPL score indicates that the language model performs better at predicting the next word. 



\begin{table}[t]
    \centering
    \tabcolsep=1mm
    \renewcommand{\arraystretch}{1.2}
    \caption{The perplexity (PPL), the p-value of the hypothesis test, and watermark success rate (WSR) of embedding a watermark into text generation models via EaaW.}
    \label{tab:nlp_wm}
    \scalebox{0.78}{
    \begin{tabular}{c|ccccccc}
    \hline
    \hline
        Dataset & Length$\rightarrow$ & No WM & 32 & 48 & 64 & 96 & 128 \\
        \hline
        \multirow{3}{*}{wikitext} & PPL & 43.33 & 46.97 & 47.88 & 48.59 & 48.78 & 51.09\\
        & p-value & / & $10^{-7}$ & $10^{-10}$ & $10^{-13}$ & $10^{-20}$ & $10^{-27}$\\
        & WSR & / & 1.000 & 1.000 & 1.000 & 1.000 & 1.000\\
        \hline
        \multirow{3}{*}{bookcorpus} & PPL & 43.75 & 44.28 & 44.76 & 45.41 & 47.52 & 49.61\\
        & p-value & / & $10^{-7}$ & $10^{-10}$ & $10^{-13}$ & $10^{-20}$ & $10^{-27}$\\
        & WSR & / & 1.000 & 1.000 & 1.000 & 1.000 & 1.000\\
        \hline
        \multirow{3}{*}{ptb-text-only} & PPL & 39.49 & 40.98 & 42.41 & 42.68 & 45.52 & 48.99\\
        & p-value & / & $10^{-7}$ & $10^{-10}$ & $10^{-13}$ & $10^{-20}$ & $10^{-27}$\\
        & WSR & / & 1.000 & 1.000 & 1.000 & 1.000 & 1.000\\
        \hline
        \multirow{3}{*}{lambada} & PPL & 42.07 & 44.21 & 44.24 & 44.48 & 44.85 & 47.99\\
        & p-value & / & $10^{-7}$ & $10^{-10}$ & $10^{-13}$ & $10^{-20}$ & $10^{-27}$\\
        & WSR & / & 1.000 & 1.000 & 1.000 & 1.000 & 1.000\\
    \hline
    \hline
    \end{tabular}
    }
    \vspace{-10pt}
\end{table}

\subsubsection{Evaluation on Effectiveness and Harmlessness} 

Table \ref{tab:nlp_wm} shows the results of applying EaaW to the text generation models. In all the experiments, the watermarks are successfully embedded into the text generation models, with a significantly low p-value and $1.0$ WSR. The results suggest the effectiveness of EaaW on the text generation models.

Table \ref{tab:nlp_wm} also indicates that the functionality of the text generation models does not significantly drop after embedding the watermark, considering that PPL is an exponential metric. Also, the longer the length of the embedded watermark, the greater the impact on the model performance. However, a longer watermark can furnish more information for verification and better security.

\begin{table*}[t]
    \centering
    \renewcommand{\arraystretch}{1.2}
    \caption{The p-value of the hypothesis test, and watermark success rate (WSR) with the watermarked model (Watermarked), independent model (Independent M.), and independent trigger (Independent T.) in text generation modeling task.}
    \label{tab:nlp_independent}
    \scalebox{0.75}{
    \begin{tabular}{c|ccccccccccc}
    \hline
    \hline
        \multirow{2}{*}{Dataset} & Length$\rightarrow$ & \multicolumn{2}{c}{32} & \multicolumn{2}{c}{48} & \multicolumn{2}{c}{64} & \multicolumn{2}{c}{96} & \multicolumn{2}{c}{128} \\
        & Scenario$\downarrow$ & p-value & WSR & p-value & WSR & p-value & WSR & p-value & WSR & p-value & WSR\\
        \hline
        \multirow{3}{*}{wikitext} & Watermarked & $10^{-7}$ & 1.000 & $10^{-10}$ & 1.000 & $10^{-13}$ & 1.000 & $10^{-20}$ & 1.000 & $10^{-27}$ & 1.000\\
        & Independent M.  & 0.217 & 0.500 & 0.308 & 0.521 & 0.301 & 0.500 & 0.657 & 0.500 & 0.745 & 0.477 \\
        & Independent T.  & 0.457 & 0.450 & 0.424 & 0.413 & 0.414 & 0.422 & 0.484 & 0.435 & 0.693 & 0.466\\
        \hline
        \multirow{3}{*}{bookcorpus} & Watermarked & $10^{-7}$ & 1.000 & $10^{-10}$ & 1.000 & $10^{-13}$ & 1.000 & $10^{-20}$ & 1.000 & $10^{-27}$ & 1.000\\
        & Independent M.  & 0.021 & 0.438 & 0.062 & 0.417 & 0.256 & 0.516 & 0.440 & 0.469 & 0.489 & 0.445 \\
        & Independent T.  & 0.296 & 0.394 & 0.565 & 0.492 & 0.355 & 0.419 & 0.725 & 0.506 & 0.520 & 0.475\\
        \hline
        \multirow{3}{*}{ptb-text-only} & Watermarked & $10^{-7}$ & 1.000 & $10^{-10}$ & 1.000 & $10^{-13}$ & 1.000 & $10^{-20}$ & 1.000 & $10^{-27}$ & 1.000\\
        & Independent M.  & 0.040 & 0.406 & 0.152 & 0.438 & 0.070 & 0.469 & 0.333 & 0.521 & 0.594 & 0.445 \\
        & Independent T.  & 0.364 & 0.381 & 0.432 & 0.475 & 0.448 & 0.541 & 0.742 & 0.490 & 0.697 & 0.503\\
        \hline
        \multirow{3}{*}{lambada} & Watermarked & $10^{-7}$ & 1.000 & $10^{-10}$ & 1.000 & $10^{-13}$ & 1.000 & $10^{-20}$ & 1.000 & $10^{-27}$ & 1.000\\
        & Independent M.  & 0.013 & 0.469 & 0.015 & 0.375 & 0.222 & 0.453 & 0.461 & 0.479 & 0.584 & 0.477 \\
        & Independent T.  & 0.284 & 0.481 & 0.351 & 0.408 & 0.254 & 0.394 & 0.634 & 0.500 & 0.602 & 0.531\\
    \hline
    \hline
    \end{tabular}
    }
    \vspace{-10pt}
\end{table*}

\subsubsection{Evaluation on Distinctiveness} 

Similar to the experiments conducted on the image classification models, we also test whether the watermark can be extracted from the independently trained language model or with independent trigger samples to validate the distinctiveness of EaaW. We also exploit the model without the watermark as the independent model and randomly choose several sequences of tokens as the independent trigger samples.

From Table \ref{tab:nlp_independent}, we can find that the p-values with the independent model or independent trigger are greater than the significant level $\alpha=0.01$, and the WSRs are around $0.5$, which is similar to the results of the experiments on the image classification model. In addition, we can find that when the length of the watermark is relatively small, e.g. $32$, the p-value with the independent model is small with a minimum of $0.013$ and close to the significant level. As the length of the watermark increases, the p-value with the independent model also increases, suggesting that embedding a watermark with more bits can obtain better security and distinctiveness.

\subsection{The Resistance to Watermark Removal Attacks}
\label{sec:robustness}

After obtaining the model from other parties, the adversaries may adopt various techniques to remove watermarks or circumvent detection. In this section, we explore whether our EaaW is resistant to them. Following the suggestions in \cite{lukas2022sok}, we consider three types of attacks, including fine-tuning attacks, model pruning attacks, and adaptive attacks.

\begin{figure}[t]
    \centering
    \includegraphics[width=0.95\linewidth]{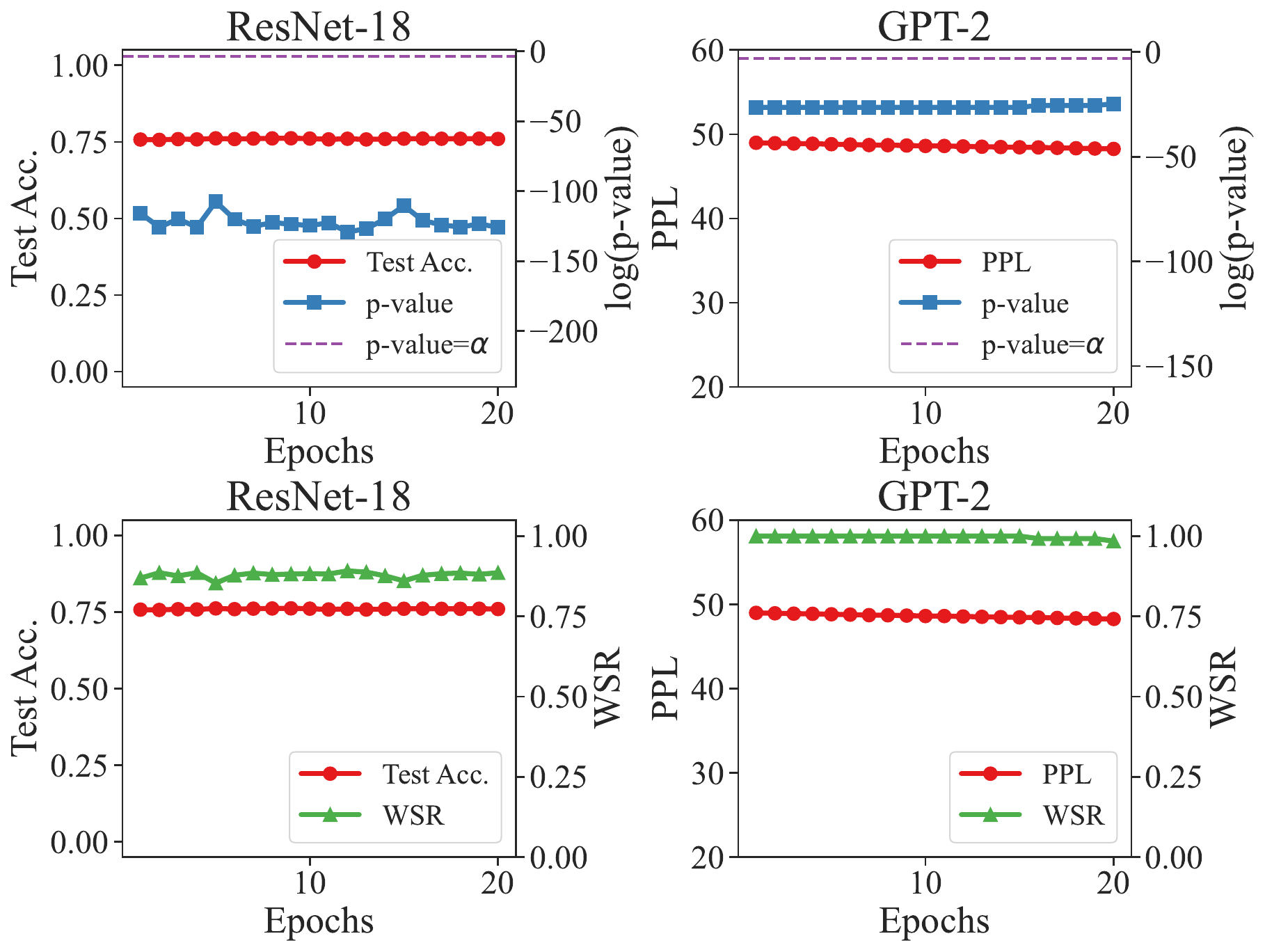}
    \caption{Watermark success rate (WSR), the log p-value, and functionality evaluation (test accuracy or PPL) of watermarked ResNet-18 and GPT-2 against fine-tuning attack.}
    \label{fig:finetuning}
    \vspace{-20pt}
\end{figure}

\begin{figure}[t]
    \centering
    \includegraphics[width=0.95\linewidth]{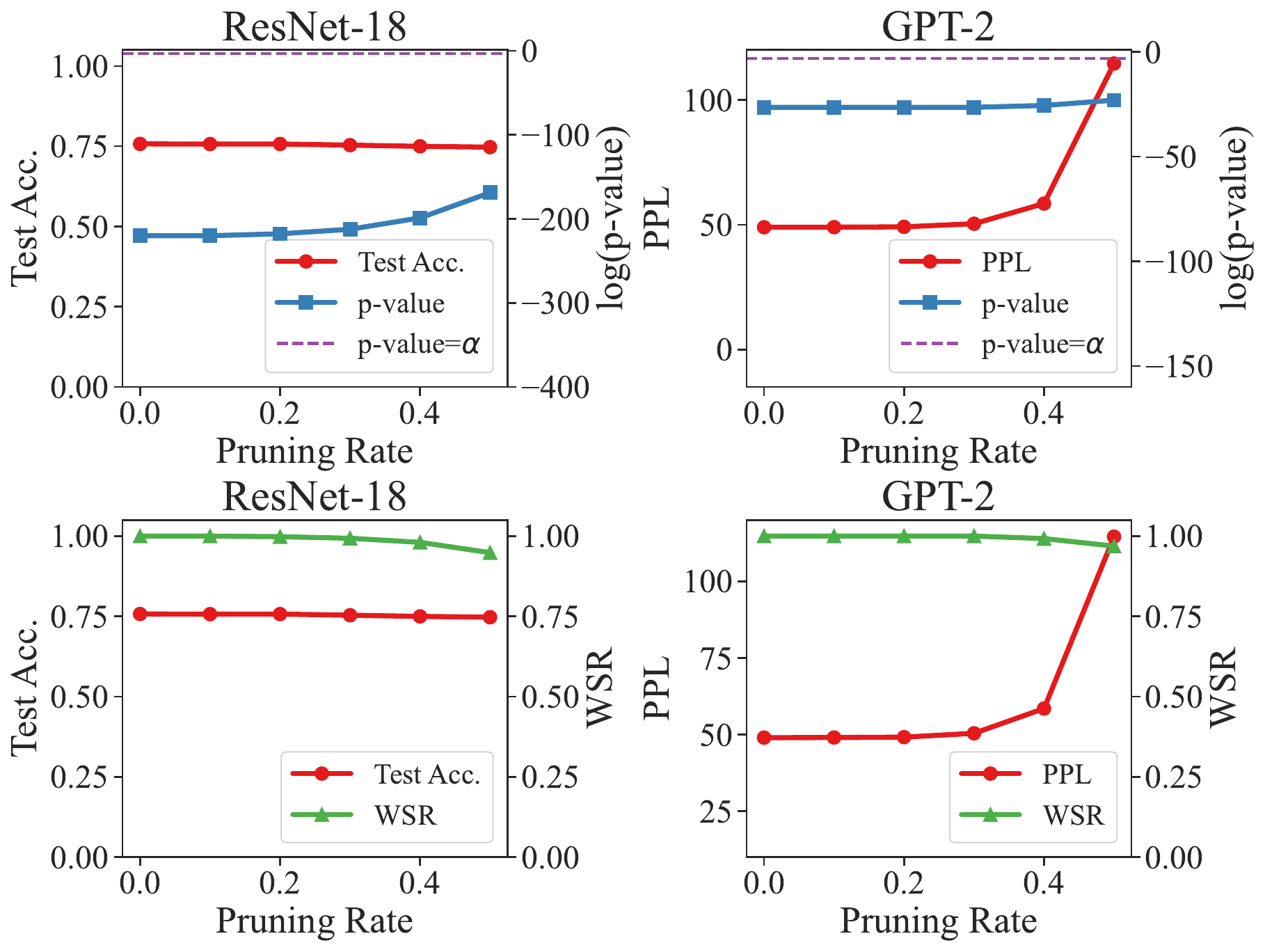}
    \caption{Watermark success rate (WSR), the log p-value, and functionality evaluation (test accuracy or PPL) of watermarked ResNet-18 and GPT-2 against model-pruning attack.}
    \label{fig:pruning}
    \vspace{-20pt}
\end{figure}

\subsubsection{The Resistance to Fine-tuning Attack} 

Fine-tuning refers to training the watermarked model with a local benign dataset for a few epochs. In the fine-tuning attack, the adversary may attempt to remove the watermark inside the model through fine-tuning. We fine-tune the EaaW-watermarked models with 20 epochs on the testing set where the training has converged.

Figure~\ref{fig:finetuning} shows the log p-value and the WSR during the fine-tuning attack. The p-value and WSR fluctuate during fine-tuning, whereas the p-value is always significantly lower than the significant level $\alpha$ (denoted by the purple dotted line) and the WSR is greater than $0.85$. These results demonstrate the resistance of our EaaW to fine-tuning attacks. We argue that this is mostly because we did not change the label of watermarked samples during model training, leading to minor effects of fine-tuning compared to backdoor-based methods.

\subsubsection{The Resistance to Model-pruning Attack} 

Model-pruning serves as a potential watermark-removal attack because it may prune watermark-related neurons. We exploit parameter pruning~\cite{han2015learning} as an example for discussion. Specifically, we prune the neurons in the model by zeroing out those with the lowest $l_1$ norm. In particular, we use the pruning rate to denote how many proportions of neurons are pruned.

As shown in Figure~\ref{fig:pruning}, the test accuracy of ResNet-18 drops while the PPL of GPT-2 increases, as the pruning rate increases. It indicates the degradation of the functionality of the model. However, the p-value of the pruned model negligibly changes and is lower than the significant level $\alpha$. Besides, the WSR is still greater than $0.9$. These results suggest that our EaaW resists the model-pruning attack.


\subsubsection{The Resistance to Adaptive Attacks}
\label{sec:adaptive}

In practice, the adversaries may know the existence of our EaaW and design adaptive attacks to circumvent it. Specifically, they may try to remove the watermark or interfere with the watermark extraction by manipulating the explanation of the input data. Existing techniques for manipulating explanations can be classified into two categories: \textbf{(1)} modifying model parameters~\cite{heo2019fooling, noppel2023disguising} or \textbf{(2)} modifying the inputs (\ie, the adversarial attack against XAI)~\cite{ghorbani2019interpretation}. In this section, we hereby present the results under the first type of attacks in two representative scenarios, namely the \emph{overwriting attack} and the \emph{unlearning attack}. The discussion on the second attack can be found in Appendix~\ref{apd:adaptive}.

\emph{Scenario 1 (Overwriting Attack):} In this scenario, we assume that the adversary knows the procedure of the EaaW method, but has no knowledge of the trigger samples and the target watermark used by the model owner. Therefore, the adversary can independently generate the trigger samples and the watermark, and then embed them into the model, attempting to overwrite the watermark embedded before. This category of adaptive attack is called the \emph{overwriting attack}.

The experimental results of the overwriting attack are shown in Table \ref{tab:adaptive1}. We conduct a $10$-epoch fine-tuning to simulate the overwriting process and the models have already converged after 10 epochs. After the overwriting attack, the functionality of both the watermarked ResNet-18 and the watermarked GPT-2 decreases, while the p-values are still low and the WSRs of the original watermarks are close to $0.9$. It indicates that the overwriting attack cannot effectively remove our EaaW watermark. We notice that the overwriting attack can embed the adversary's watermark into the victim model to some extent. As shown in Table~\ref{tab:adaptive1}, the WSRs of the new watermarks are larger than $0.8$, indicating that there are two distinct watermarks within the model. However, this is a common and trivial situation in watermarking. This issue can be easily solved by registering the watermark and the model to a trusted third party (\eg, the intellectual property office) accompanied by timestamps~\cite{waheed2024grove}. The watermark with a later timestamp will not be treated as a valid copyright certificate.


\begin{table}[t]
    \tabcolsep=1mm
    \renewcommand{\arraystretch}{1.1}
    \centering
    \caption{Watermark success rate (WSR) of the original watermark (dubbed `Ori. WM') and the adversary's new watermark (dubbed `New WM'), the log p-value, and functionality evaluation (test accuracy or PPL) of ResNet-18 and GPT-2 against overwriting attack and unlearning attack.}
    \label{tab:adaptive1}
    \scalebox{0.76}{
    \begin{tabular}{c|cccc}
    \hline
    \hline
        Model$\downarrow$ & Metric$\downarrow$ & Before & After Overwriting & After Unlearning \\
        \hline
        \multirow{4}{*}{ResNet-18} & Test Acc. & 75.72 & 69.18 & 73.62\\
        & p-value & $10^{-222}$ & $10^{-134}$ & $10^{-127}$\\
        & WSR of Ori. WM & 1.000 & 0.899 & 0.888\\
        & WSR of New WM & / & 0.815 & / \\
        \hline
        \multirow{4}{*}{GPT-2} & PPL & 48.99 & 50.29 & 48.96\\
        & p-value & $10^{-27}$ & $10^{-18}$ & $10^{-24}$\\
        & WSR of Ori. WM & 1.000 & 0.906 & 0.969\\
        & WSR of New WM & / & 0.883 & / \\
         \hline
         \hline
    \end{tabular}
    }
    \vspace{-10pt}
\end{table}



\emph{Scenario 2 (Unlearning Attack):} In this scenario, we assume that the adversary knows the embedded watermark, but still has no knowledge of the trigger samples. As such, the adversary will randomly select some trigger samples and try to unlearn the watermark by updating the model in the direction opposite to the watermarking gradient. We make this assumption because the target watermark can often be conjectured. For example, the watermark may be the logo of the corporation or the profile photo of the individual developer. This type of attack is called the \emph{unlearning attack}. The adversary uses the following loss function to unlearn the watermark:
\begin{equation}
    \label{eq:adaptive2}
    \left.
    \begin{aligned}
        \min_{\hat{\Theta}}\mathcal{L}_1(f(\mathcal{X}, \hat{\Theta}), \mathcal{Y}) - r_1 \mathcal{L}_2({\tt explain}(\tilde{\mathcal{X}}_T, \tilde{\mathcal{Y}}_T, \hat{\Theta}),\bm{\mathcal{W}}).
    \end{aligned}
    \right.
\end{equation}

The experimental results of the unlearning attack are illustrated in Table \ref{tab:adaptive1}. The results demonstrate that our EaaW also resists unlearning attacks. Specifically, the WSRs drop only $0.112$ and $0.031$, respectively. The watermark can still be extracted from the model and the ownership can still be verified with low p-values. 


\subsection{Ablation Study}
\label{sec:ablation}

In this section, we conduct the ablation study to investigate the effect of some important hyper-parameters used in EaaW, such as the size of the trigger samples, the number $c$ of the masks, and the coefficient $r_1$. More ablation studies can be found in Appendix~\ref{apd:ablation}.

\subsubsection{Effect of the Size of the Trigger Samples} 

\begin{figure}[t]
    \centering
    \includegraphics[width=0.95\linewidth]{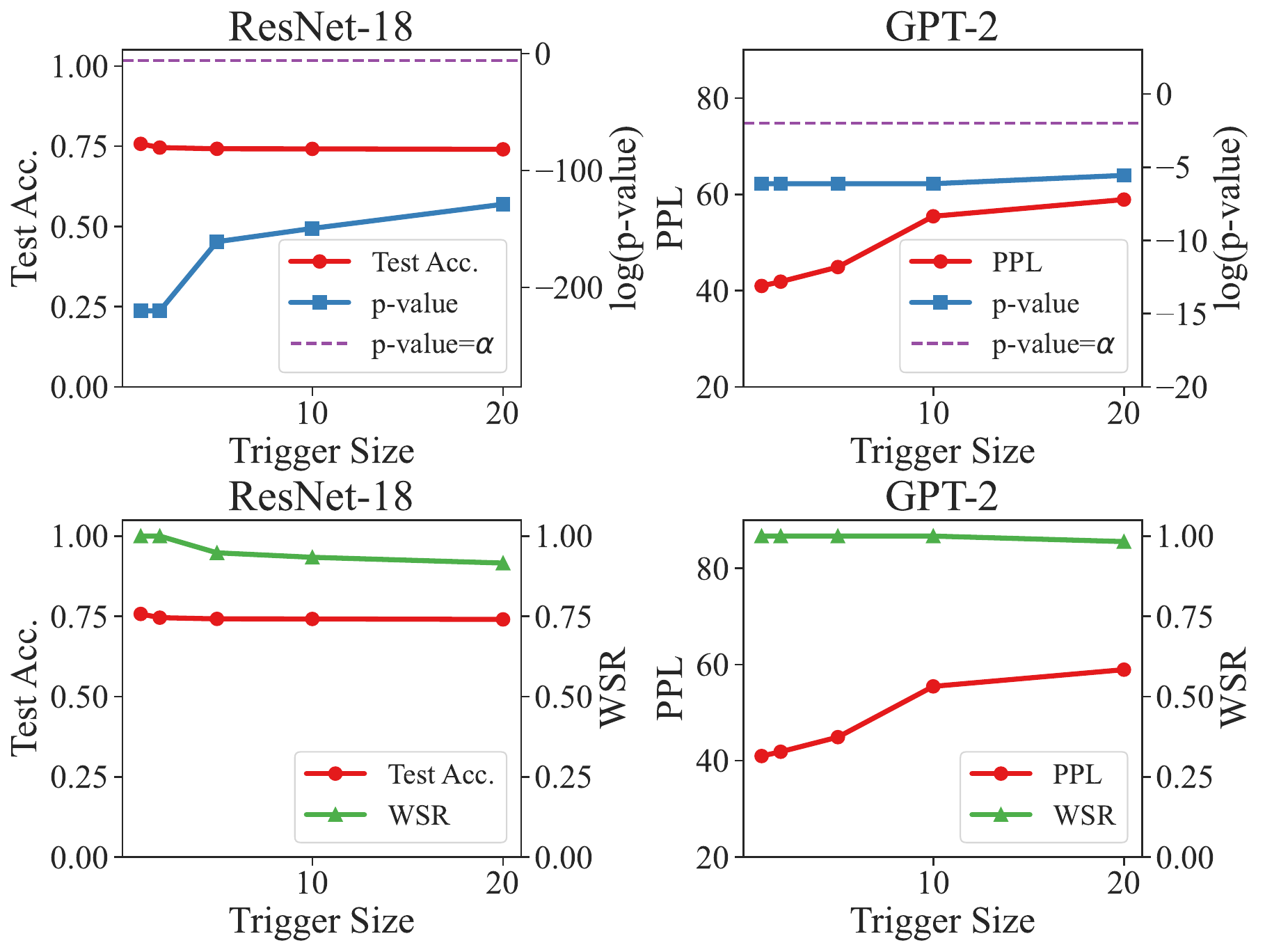}
    \caption{The Watermark success rate (WSR), the log p-value, and the functionality evaluation metrics (test accuracy or PPL) of watermarked ResNet-18 and GPT-2 with different sizes of the trigger samples.}
    \label{fig:trigger_size}
    \vspace{-20pt}
\end{figure}

In EaaW, the trigger sample $\mathcal{X}_T$ and its label $\mathcal{Y}_T$ can be considered as the secret key to extracting the watermark. Holding one secret key is enough for most cases, while multiple secret keys can further enhance the security of EaaW. In this section, we select $1$, $2$, $5$, $10$, and $20$ trigger samples and test the effectiveness of EaaW with different numbers of trigger samples. The results are illustrated in Figure~\ref{fig:trigger_size}. As the size of the trigger samples increases, 
the functionality of the watermarked model and the WSR degrades, and the p-value increases. But generally speaking, the model is capable of accommodating multiple trigger samples and the model owner can choose the size of trigger samples based on its practical requirements.

\begin{figure}[t]
    \centering
    \includegraphics[width=0.95\linewidth]{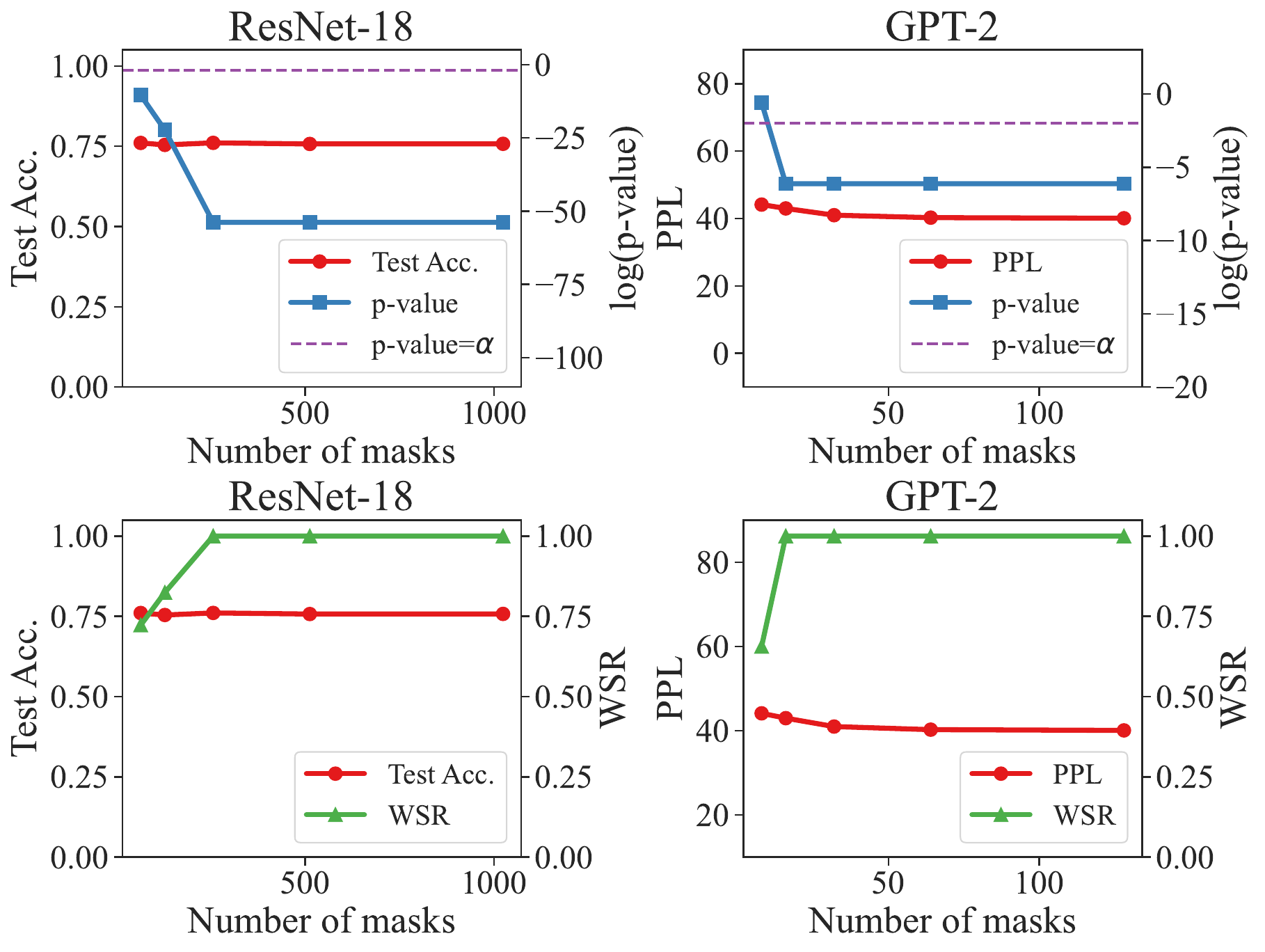}
    \caption{The watermark success rate (WSR), the log p-value, and the functionality evaluation metrics (test accuracy or PPL) of ResNet-18 and GPT-2 with different numbers of masks.}
    \label{fig:mask}
    \vspace{-20pt}
\end{figure}

\begin{figure}[t]
    \centering
    \includegraphics[width=0.95\linewidth]{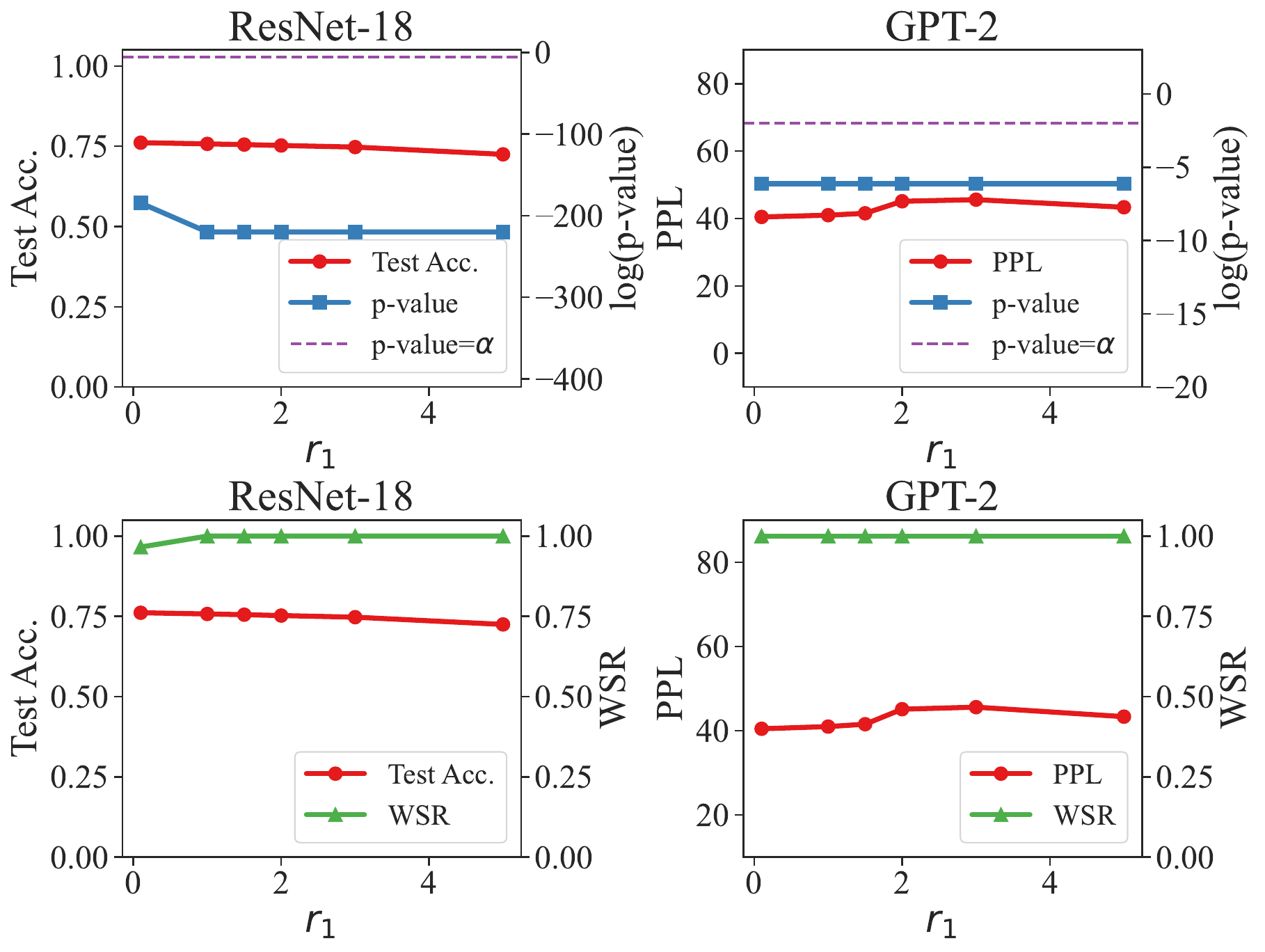}
    \caption{The watermark success rate (WSR), the log p-value, and the functionality evaluation (test accuracy or PPL) of watermarked ResNet-18 and GPT-2 with different $r_1$.}
    \label{fig:r1}
    \vspace{-20pt}
\end{figure}

\subsubsection{Effect of the Number of the Masks}

In this section, we study the effect of the number $c$ of the masks and masked samples. We set $c$ to be $64$ to $1024$ to embed a $256$-bit watermark into ResNet-18 and set $c$ to be $8$ to $128$ to embed a $32$-bit watermark into GPT-2. The results are in Figure~\ref{fig:mask}. The results indicate that using a low $c$ may lead to the failure of embedding the watermark. A small number of masked samples can not effectively evaluate the importance of each basic part, and thus the feature attribution algorithm does not work well in that case. On the contrary, sampling more masked samples contributes to the extraction of the watermark and can better preserve the functionality of the watermarked model. However, the overhead of embedding and extracting the watermark also increases. There is a trade-off between functionality and efficiency.

\subsubsection{Effect of Coefficient $r_1$} 

$r_1$ is the coefficient of the watermark loss in Eq.~(\ref{eq:wm}). $r_1$ governs the balance between embedding the watermark and preserving model functionality. To assess its impact, we varied $r_1$ across values of 0.1, 1.0, 1.5, 2.0, 3.0, and 5.0 for evaluation purposes. Figure~\ref{fig:r1} reveals that employing a larger $r_1$ can potentially exert a more pronounced negative effect on the functionality of the watermarked model; conversely, adopting a smaller $r_1$ may not entirely ensure complete watermark embedding within the model's framework. Nevertheless, in most scenarios, successful watermark integration into the model was achieved overall.

\section{Discussion and Analysis}
\label{sec:discussion}

\begin{table*}[t]
    \centering
    \tabcolsep=1.15mm
    \renewcommand{\arraystretch}{1.15}
    \caption{The watermark success rate (WSR), the harmless degree $H$ (larger is better), and test accuracy (Test Acc.) using the backdoor-based model watermarking method and EaaW in the image classification task.}
    \label{tab:backdoor}
    \scalebox{0.77}{
    \begin{tabular}{c|cc|ccccccccccccccc}
    \hline
    \hline
        \multirow{2}{*}{Dataset} & \multirow{2}{*}{\makecell[c]{Length /\\ Trigger Size}} & Trigger$\rightarrow$ & \multicolumn{3}{c}{Noise~\cite{liu2021secure}} & \multicolumn{3}{c}{Unrelated~\cite{zhang2018protecting}} & \multicolumn{3}{c}{Mask~\cite{guo2018watermarking}} & \multicolumn{3}{c}{Patch~\cite{zhang2018protecting}} & \multicolumn{3}{c}{Black-edge}\\
        & & Method$\downarrow$ & Test Acc. & $H$ & WSR & Test Acc. & $H$ & WSR & Test Acc. & $H$ & WSR & Test Acc. & $H$ & WSR & Test Acc. & $H$ & WSR \\
        \hline
        \multirow{9}{*}{CIFAR-10} & \multirow{3}{*}{64} & No WM & 90.54 & / & / & 90.54 & / & / & 90.54 & / & / & 90.54 & / & /  & 90.54 & / & /\\
        & & Backdoor & 90.38 & 89.74 & \textbf{1.000} & 88.74 & 88.10 & \textbf{1.000} & 90.34 & 89.71 & 0.984 & 84.28 & 83.64 & \textbf{1.000} & 86.24 & 85.60 & \textbf{1.000}\\
        & & EaaW & \textbf{90.49} & \textbf{90.48} & \textbf{1.000} & \textbf{90.49} & \textbf{90.48} & \textbf{1.000} & \textbf{90.46} & \textbf{90.47} & \textbf{1.000} & \textbf{90.38} & \textbf{90.39} & \textbf{1.000} & \textbf{90.37} & \textbf{90.38} & \textbf{1.000}\\
        \cline{2-18}
        & \multirow{3}{*}{256} & No WM & 90.54 & / & / & 90.54 & / & / & 90.54 & / & / & 90.54 & / & /  & 90.54 & / & / \\
        & & Backdoor & 90.33 & 87.77 & \textbf{1.000} & 87.99 & 85.43 & \textbf{1.000} & 90.28 & 87.72 & \textbf{1.000} & \textbf{90.11} & 87.75 & \textbf{1.000} & 90.07 & 87.51 & \textbf{1.000}\\
        & & EaaW & \textbf{90.53} & \textbf{90.52} & \textbf{1.000} & \textbf{90.28} & \textbf{90.27} & \textbf{1.000} & \textbf{90.49} & \textbf{90.50} & \textbf{1.000} & \textbf{90.11} & \textbf{90.12} & \textbf{1.000} & \textbf{90.35} & \textbf{90.36} & \textbf{1.000}\\
        \cline{2-18}
        & \multirow{3}{*}{1024} & No WM & 90.54 & / & /  & 90.54 & / & /  & 90.54 & / & /  & 90.54 & / & /  & 90.54 & / & / \\
        & & Backdoor & 90.19 & 80.19 & 0.977 & 88.14 & 77.93 & 0.997 & 90.17 & 79.93 & \textbf{1.000} & \textbf{90.03} & 79.79 & \textbf{1.000} & \textbf{89.81} & 79.57 & \textbf{1.000}\\
        & & EaaW & \textbf{90.39} & \textbf{90.38} & \textbf{1.000} & \textbf{90.01} & \textbf{90.00} & 0.989 & \textbf{90.38} & \textbf{90.39} & \textbf{1.000} & 89.04 & \textbf{89.05} & 0.998 & 89.04 & \textbf{89.05} & \textbf{1.000}\\
        \hline
        \multirow{9}{*}{ImageNet} & \multirow{3}{*}{64} & No WM & 76.38 & / & / & 76.38 & / & / & 76.38 & / & / & 76.38 & / & / & 76.38 & / & /\\
        & & Backdoor & 73.16 & 72.67 & 0.766 & 75.94 & 75.30 & \textbf{1.000} & 75.06 & 74.42 & \textbf{1.000} & 74.18 & 73.54 & \textbf{1.000} & 73.96 & 73.32 & \textbf{1.000}\\
        & & EaaW & \textbf{75.80} & \textbf{75.79} & \textbf{1.000} & \textbf{76.00} & \textbf{75.99} & \textbf{1.000} & \textbf{75.98} & \textbf{75.99} & \textbf{1.000} & \textbf{75.76} & \textbf{75.77} & \textbf{1.000} & \textbf{75.78} & \textbf{75.79} & \textbf{1.000}\\
        \cline{2-18}
        & \multirow{3}{*}{256} & No WM & 76.38 & / & /  & 76.38 & / & /  & 76.38 & / & /  & 76.38 & / & /   & 76.38 & / & /  \\
        & & Backdoor & 73.70 & 71.14 & \textbf{1.000} & 75.92 & 73.36 & \textbf{1.000} & 74.08 & 71.52 & \textbf{1.000} & 70.34 & 67.80 & 0.992 & 71.10 & 68.59 & 0.980\\
        & & EaaW & \textbf{75.86} & \textbf{75.85} & \textbf{1.000} & \textbf{76.36} & \textbf{76.35} & \textbf{1.000} & \textbf{76.06} & \textbf{76.07}  & \textbf{1.000} & \textbf{76.06} & \textbf{76.07} & \textbf{1.000} & \textbf{75.60} & \textbf{75.61} & \textbf{1.000}\\
        \cline{2-18}
        & \multirow{3}{*}{1024} & No WM & 76.38 & / & /  & 76.38 & / & /  & 76.38 & / & /  & 76.38 & / & /   & 76.38 & / & /  \\
        & & Backdoor & 73.56 & 64.22 & 0.912 & \textbf{75.86} & 65.62 & \textbf{1.000} & 74.86 & 64.62 & \textbf{1.000} & 73.92 & 63.68 & \textbf{1.000} & \textbf{74.32} & 64.08 & \textbf{1.000}\\
        & & EaaW & \textbf{75.40} & \textbf{75.39} & \textbf{1.000} & 75.26 & \textbf{75.25} & 0.999 & \textbf{75.74} & \textbf{75.75} & \textbf{1.000} & \textbf{73.48} & \textbf{73.49} & 0.999 & 72.84 & \textbf{72.85} & \textbf{1.000}\\
         \hline
         \hline
    \end{tabular}
    }
    \vspace{-10pt}
\end{table*}

\subsection{How EaaW Affect the Watermarked Model?}

\begin{figure}[t]
    \centering
    \includegraphics[width=0.95\linewidth]{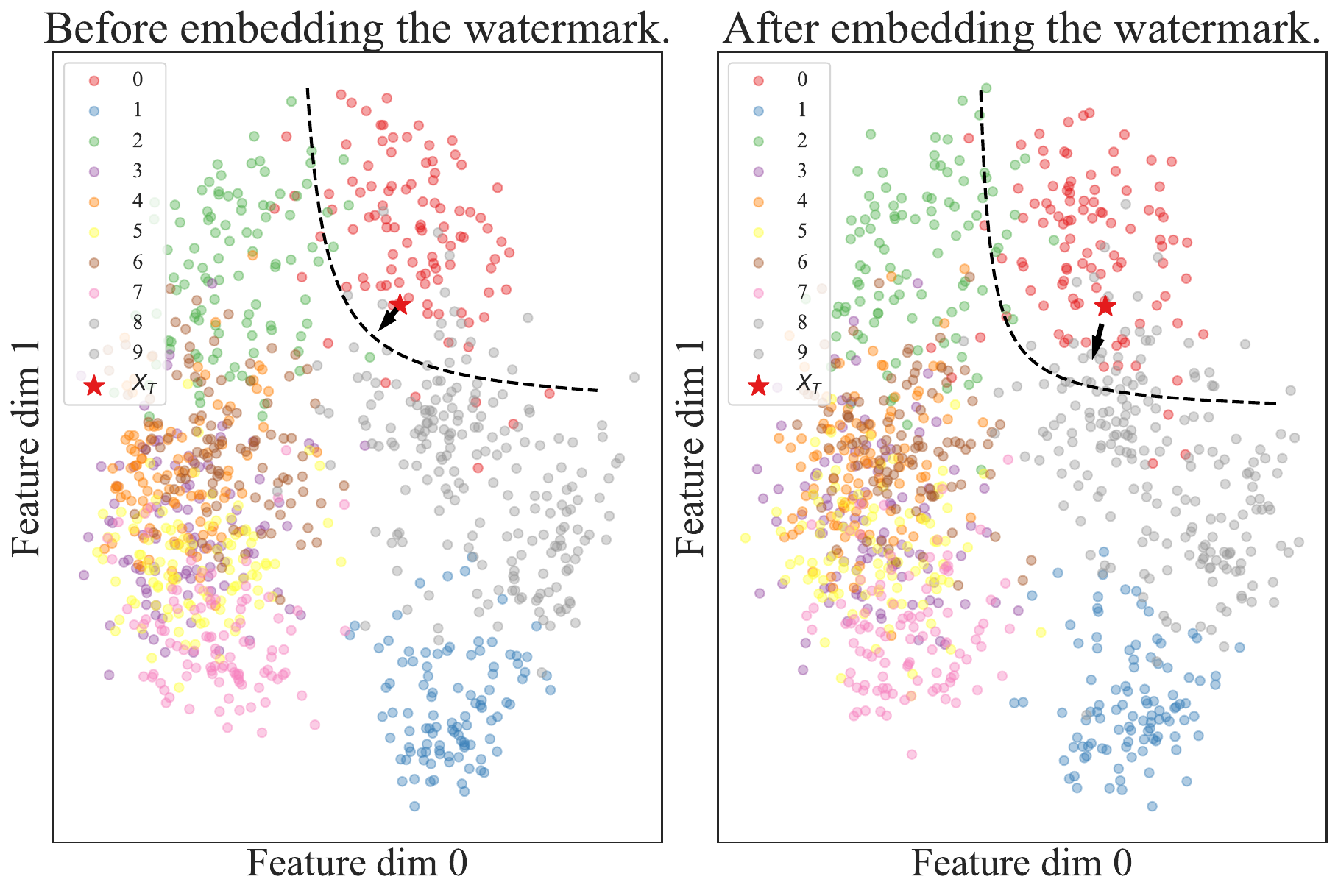}
    \caption{The visualization of the feature representations of the training data before and after embedding the watermark. Those of the trigger sample is marked as `star' and the trigger sample belongs to the class $0$ colored by red. The feature representations of all the samples do not significantly change. Instead, the direction of the trigger sample to the decision boundary is transformed. (Better viewed in color)}
    \label{fig:visual}
    \vspace{-15pt}
\end{figure}

In this section, we explore how EaaW affects the watermarked model. Inspired by~\cite{hua2023deep}, we analyze the effect of EaaW by visualizing the intermediate features of both the benign samples and the trigger sample. We first randomly select $100$ images per class from the training dataset. Subsequently, we input those data into the model and get the output of the features by the penultimate layer. To further analyze these features, we employ the kernel principal component analysis (Kernel PCA) algorithm for dimensionality reduction, reducing those features to $2$ dimensions. The visualization of these reduced features is presented in Figure~\ref{fig:visual}. Notably, the feature corresponding to the trigger sample is denoted as a star.

From Figure~\ref{fig:visual}, we can see that the intermediate feature representations are generally unaffected before and after embedding the watermark using EaaW. The feature of the trigger sample continues to reside within the cluster corresponding to its respective class (the red class numbered $0$), thereby demonstrating the harmlessness of EaaW. Additionally, as depicted in Figure~\ref{fig:visual}, EaaW actually changes the direction of the trigger sample to the decision boundary of the model, that is, the direction of the trigger sample's gradient (as visualized as the black arrows towards the general decision boundary). By embedding a multi-bit watermark into the sign of this gradient, EaaW achieves successful model watermarking without altering the prediction of the trigger sample.

\subsection{Security Analysis against Ambiguity Attack}
\label{sec:ambiguity}

When the adversary acquires the watermarked model, 
the adversary can attempt to forge a fake watermark to establish its ownership of the watermarked model. If both the true model owner and the adversary can independently authenticate their copyright claims, it becomes impossible to ascertain the actual ownership. This type of attack is named \emph{ambiguity attack}~\cite{fan2019rethinking} or false claim attack~\cite{liu2023false}. More details are in Appendix~\ref{apd:falseclaim}. The ambiguity attack for our EaaW is as follows.

\begin{definition}[Ambiguity Attack]
    \label{def:ambiguity}
    Given a watermarked model $\hat{\Theta}$, the objective of the ambiguity attack is to forge a fake trigger sample $\tilde{\mathcal{X}}_T, \tilde{\mathcal{Y}}_T$ that can be utilized to extract its own watermark $\tilde{\mathcal{W}}$ and pass the ownership verification algorithm described in Algorithm \ref{algo:ov}.
\end{definition}
As our approach involves embedding a multi-bit watermark into the model, the EaaW technique offers superior security compared to the zero-bit backdoor-based model watermarking methods. Intuitively, assuming that the probability of a successful ambiguity attack against a one-bit (or zero-bit) watermark method is $1/\xi$, then the probability of a successful ambiguity attack against a $k$-bit watermark method is $1/\xi^k$. As such, for a not-too-small length $k$ of the watermark, our proposed EaaW can withstand ambiguity attacks. To support this claim, we present the following proposition.

\begin{proposition}
    \label{prop:ambiguity}
    Given the length of the watermark $k$, the probability of a successful ambiguity attack is $1/2^k$.
\end{proposition}

\begin{proof}
    Assuming that the adversary has no knowledge of the trigger sample used for ownership verification. When the adversary tries to forge a trigger sample by random selection, the probability of each bit being the correct bit can be assumed as $1/2$. Since the explanation output by the feature attribution algorithm depends on all the features of the input data, the probability of the explanation correctly matching the watermark is $1/2^k$.
\end{proof}

Proposition~\ref{prop:ambiguity} shows that the time complexity of the ambiguity attack against EaaW is exponential concerning the length $k$ of the watermark, indicating that our watermarking method is hard to forge by the adversary and is resistant to ambiguity attack. Furthermore, Proposition~\ref{prop:ambiguity} also suggests utilizing a multi-bit watermark can obtain better security.




\subsection{The Comparision to Backdoor Watermarks}
\label{sec:compare}

Backdoor-based model watermarks are currently the most representative and popular black-box model ownership verification techniques. Arguably, the primary differences among these various backdoor-based approaches lie in their distinct construction of the trigger set~\cite{sun2023deep}. 
Recall that existing literature has already shed light on the ambiguous nature of backdoor-based watermarks~\cite{liu2023false, hua2023unambiguous}, while our Section \ref{sec:ambiguity} demonstrates that our EaaW technique remains resilient against ambiguity attacks. Accordingly, this section primarily focuses on the comparison between various backdoor-based methods~\cite{zhang2018protecting, guo2018watermarking, liu2021secure} and our EaaW in terms of effectiveness and harmlessness. Inspired by the definition in~\cite{guo2023domain}, we define the \emph{harmless degree} $H$ as the metric for evaluating the level of harmlessness. $H$ is defined as the accuracy achieved both on the benign testing dataset $\mathcal{X},\mathcal{Y}$ and the trigger set $\mathcal{X}_T, \mathcal{Y}_T$ with the ground-truth labels, as follows. 
\begin{equation}
    H=\frac{1}{|\mathcal{X}\cup\mathcal{X}_T|}\sum_{\bm{x}\in \mathcal{X}\cup\mathcal{X}_T}\mathbb{I}\{f(\bm{x};\Theta)=g(\bm{x})\},
\end{equation}
where $\mathbb{I}\{\cdot\}$ is the indicator function and $g(\bm{x})$ always output the ground-truth label of $\bm{x}$. A larger $H$ means the watermarks have less effect on the utility of the models.

As shown in Table~\ref{tab:backdoor}, the watermarking effectiveness of our EaaW is on par with or even better than that of the baseline backdoor-based methods. Regarding harmlessness, our EaaW approach outperforms the backdoor-based techniques as evidenced by higher harmless degrees $H$. For example, the harmlessness degree $H$ of our EaaW is nearly 10\% higher than that of backdoor-based watermarks in all cases with trigger size 1024. Note that backdoor-based watermarks introduce backdoors that can be exploited by adversaries to generate specific malicious predictions, although they do not compromise performance on benign samples.




\section{Conclusion}
\label{sec:conclusion}

In this paper, we revealed that the widely applied backdoor-based model watermarking methods have two major drawbacks, including harmlessness and ambiguity. 
We found out that those limitations can both be attributed to that the backdoor-based watermark utilizes misclassification to embed a `zero-bit' watermark into the model. To tackle these issues, we proposed a harmless and multi-bit model ownership verification method, named Explanation as a Watermark (EaaW), inspired by XAI. EaaW is the first to introduce the insight of embedding the multi-bit watermark into the explanation output by feature attribution methods. 
We correspondingly designed a feature attribution-based watermark embedding and extraction algorithm. Our empirical experiments demonstrated the effectiveness, distinctiveness, and harmlessness of EaaW. We hope our EaaW can provide a new angle and deeper understanding of model ownership verification to facilitate secure and trustworthy model deployment and sharing.



\section*{Acknowledgment}

This research is supported in part by the National Key Research and Development Program of China under Grant 2021YFB3100300 and the National Natural Science Foundation of China under Grants (62072395 and U20A20178). This work was mostly done when Yiming Li was at the State Key Laboratory of Blockchain and Data Security, Zhejiang University. He is currently at Nanyang Technological University. 




\bibliographystyle{IEEEtranS}
\bibliography{IEEEabrv,ref}

\begin{thebibliography}{10}
\providecommand{\url}[1]{#1}
\csname url@samestyle\endcsname
\providecommand{\newblock}{\relax}
\providecommand{\bibinfo}[2]{#2}
\providecommand{\BIBentrySTDinterwordspacing}{\spaceskip=0pt\relax}
\providecommand{\BIBentryALTinterwordstretchfactor}{4}
\providecommand{\BIBentryALTinterwordspacing}{\spaceskip=\fontdimen2\font plus
\BIBentryALTinterwordstretchfactor\fontdimen3\font minus
  \fontdimen4\font\relax}
\providecommand{\BIBforeignlanguage}[2]{{%
\expandafter\ifx\csname l@#1\endcsname\relax
\typeout{** WARNING: IEEEtranS.bst: No hyphenation pattern has been}%
\typeout{** loaded for the language `#1'. Using the pattern for}%
\typeout{** the default language instead.}%
\else
\language=\csname l@#1\endcsname
\fi
#2}}
\providecommand{\BIBdecl}{\relax}
\BIBdecl

\bibitem{adi2018turning}
Y.~Adi, C.~Baum, M.~Cisse, B.~Pinkas, and J.~Keshet, ``Turning your weakness
  into a strength: Watermarking deep neural networks by backdooring,'' in
  \emph{USENIX Security}, 2018.

\bibitem{brown2020language}
T.~Brown, B.~Mann, N.~Ryder, M.~Subbiah, J.~D. Kaplan, P.~Dhariwal,
  A.~Neelakantan, P.~Shyam, G.~Sastry, A.~Askell \emph{et~al.}, ``Language
  models are few-shot learners,'' in \emph{NeurIPS}, 2020.

\bibitem{cao2021ipguard}
X.~Cao, J.~Jia, and N.~Z. Gong, ``Ipguard: Protecting intellectual property of
  deep neural networks via fingerprinting the classification boundary,'' in
  \emph{AsiaCCS}, 2021.

\bibitem{chen2019blackmarks}
H.~Chen, B.~D. Rouhani, and F.~Koushanfar, ``Blackmarks: Blackbox multibit
  watermarking for deep neural networks,'' \emph{arXiv preprint
  arXiv:1904.00344}, 2019.

\bibitem{chen2022copy}
J.~Chen, J.~Wang, T.~Peng, Y.~Sun, P.~Cheng, S.~Ji, X.~Ma, B.~Li, and D.~Song,
  ``Copy, right? a testing framework for copyright protection of deep learning
  models,'' in \emph{S\&P}, 2022.

\bibitem{cong2022sslguard}
T.~Cong, X.~He, and Y.~Zhang, ``Sslguard: A watermarking scheme for
  self-supervised learning pre-trained encoders,'' in \emph{CCS}, 2022.

\bibitem{deng2009imagenet}
J.~Deng, W.~Dong, R.~Socher, L.-J. Li, K.~Li, and L.~Fei-Fei, ``Imagenet: A
  large-scale hierarchical image database,'' in \emph{CVPR}, 2009.

\bibitem{devlin2018bert}
J.~Devlin, M.-W. Chang, K.~Lee, and K.~Toutanova, ``Bert: Pre-training of deep
  bidirectional transformers for language understanding,'' \emph{arXiv preprint
  arXiv:1810.04805}, 2018.

\bibitem{dosovitskiy2020image}
A.~Dosovitskiy, L.~Beyer, A.~Kolesnikov, D.~Weissenborn, X.~Zhai,
  T.~Unterthiner, M.~Dehghani, M.~Minderer, G.~Heigold, S.~Gelly \emph{et~al.},
  ``An image is worth 16x16 words: Transformers for image recognition at
  scale,'' in \emph{ICLR}, 2020.

\bibitem{dwivedi2023explainable}
R.~Dwivedi, D.~Dave, H.~Naik, S.~Singhal, R.~Omer, P.~Patel, B.~Qian, Z.~Wen,
  T.~Shah, G.~Morgan \emph{et~al.}, ``Explainable ai (xai): Core ideas,
  techniques, and solutions,'' \emph{ACM Computing Surveys}, vol.~55, no.~9,
  pp. 1--33, 2023.

\bibitem{dziedzic2022dataset}
A.~Dziedzic, H.~Duan, M.~A. Kaleem, N.~Dhawan, J.~Guan, Y.~Cattan, F.~Boenisch,
  and N.~Papernot, ``Dataset inference for self-supervised models,'' in
  \emph{NeurIPS}, 2022.

\bibitem{fan2019rethinking}
L.~Fan, K.~W. Ng, and C.~S. Chan, ``Rethinking deep neural network ownership
  verification: Embedding passports to defeat ambiguity attacks,'' in
  \emph{NeurIPS}, 2019.

\bibitem{ghorbani2019interpretation}
A.~Ghorbani, A.~Abid, and J.~Zou, ``Interpretation of neural networks is
  fragile,'' in \emph{AAAI}, 2019.

\bibitem{gu2019badnets}
T.~Gu, K.~Liu, B.~Dolan-Gavitt, and S.~Garg, ``Badnets: Evaluating backdooring
  attacks on deep neural networks,'' \emph{IEEE Access}, vol.~7, pp.
  47\,230--47\,244, 2019.

\bibitem{guo2018watermarking}
J.~Guo and M.~Potkonjak, ``Watermarking deep neural networks for embedded
  systems,'' in \emph{ICCAD}, 2018.

\bibitem{guo2023domain}
J.~Guo, Y.~Li, L.~Wang, S.-T. Xia, H.~Huang, C.~Liu, and B.~Li, ``Domain
  watermark: Effective and harmless dataset copyright protection is closed at
  hand,'' in \emph{NeurIPS}, 2023.

\bibitem{han2015learning}
S.~Han, J.~Pool, J.~Tran, and W.~Dally, ``Learning both weights and connections
  for efficient neural network,'' in \emph{NeurIPS}, 2015.

\bibitem{he2016deep}
K.~He, X.~Zhang, S.~Ren, and J.~Sun, ``Deep residual learning for image
  recognition,'' in \emph{CVPR}, 2016.

\bibitem{he2023finer}
Y.~He, J.~Lou, Z.~Qin, and K.~Ren, ``Finer: Enhancing state-of-the-art
  classifiers with feature attribution to facilitate security analysis,'' in
  \emph{CCS}, 2023.

\bibitem{heo2019fooling}
J.~Heo, S.~Joo, and T.~Moon, ``Fooling neural network interpretations via
  adversarial model manipulation,'' in \emph{NeurIPS}, 2019.

\bibitem{hua2023deep}
G.~Hua and A.~B.~J. Teoh, ``Deep fidelity in dnn watermarking: A study of
  backdoor watermarking for classification models,'' \emph{Pattern
  Recognition}, vol. 144, 2023.

\bibitem{hua2023unambiguous}
G.~Hua, A.~B.~J. Teoh, Y.~Xiang, and H.~Jiang, ``Unambiguous and high-fidelity
  backdoor watermarking for deep neural networks,'' \emph{IEEE Transactions on
  Neural Networks and Learning Systems}, 2023.

\bibitem{jia2021zest}
H.~Jia, H.~Chen, J.~Guan, A.~S. Shamsabadi, and N.~Papernot, ``A zest of lime:
  Towards architecture-independent model distances,'' in \emph{ICLR}, 2021.

\bibitem{jia2021entangled}
H.~Jia, C.~A. Choquette-Choo, V.~Chandrasekaran, and N.~Papernot, ``Entangled
  watermarks as a defense against model extraction,'' in \emph{USENIX
  Security}, 2021.

\bibitem{krauss2024automatic}
T.~Krau{\ss}, J.~K{\"o}nig, A.~Dmitrienko, and C.~Kanzow, ``Automatic
  adversarial adaption for stealthy poisoning attacks in federated learning,''
  in \emph{NDSS}, 2024.

\bibitem{krauss2024clearstamp}
T.~Krau{\ss}, J.~Stang, and A.~Dmitrienko, ``Clearstamp: A human-visible and
  robust model-ownership proof based on transposed model training,'' in
  \emph{USENIX Security}, 2024.

\bibitem{krizhevsky2009learning}
A.~Krizhevsky, G.~Hinton \emph{et~al.}, ``Learning multiple layers of features
  from tiny images,'' \emph{Master's thesis, University of Tront}, 2009.

\bibitem{li2023plmmark}
P.~Li, P.~Cheng, F.~Li, W.~Du, H.~Zhao, and G.~Liu, ``Plmmark: A secure and
  robust black-box watermarking framework for pre-trained language models,'' in
  \emph{AAAI}, 2023.

\bibitem{li2022untargeted}
Y.~Li, Y.~Bai, Y.~Jiang, Y.~Yang, S.-T. Xia, and B.~Li, ``Untargeted backdoor
  watermark: Towards harmless and stealthy dataset copyright protection,'' in
  \emph{NeurIPS}, 2022.

\bibitem{li2022backdoor}
Y.~Li, Y.~Jiang, Z.~Li, and S.-T. Xia, ``Backdoor learning: A survey,''
  \emph{IEEE Transactions on Neural Networks and Learning Systems}, 2022.

\bibitem{li2022move}
Y.~Li, L.~Zhu, X.~Jia, Y.~Bai, Y.~Jiang, S.-T. Xia, and X.~Cao, ``Move:
  Effective and harmless ownership verification via embedded external
  features,'' \emph{arXiv preprint arXiv:2208.02820}, 2022.

\bibitem{li2022defending}
Y.~Li, L.~Zhu, X.~Jia, Y.~Jiang, S.-T. Xia, and X.~Cao, ``Defending against
  model stealing via verifying embedded external features,'' in \emph{AAAI},
  2022.

\bibitem{li2023protecting}
Z.~Li, C.~Wang, S.~Wang, and C.~Gao, ``Protecting intellectual property of
  large language model-based code generation apis via watermarks,'' in
  \emph{CCS}, 2023.

\bibitem{lim2022protect}
J.~H. Lim, C.~S. Chan, K.~W. Ng, L.~Fan, and Q.~Yang, ``Protect, show, attend
  and tell: Empowering image captioning models with ownership protection,''
  \emph{Pattern Recognition}, vol. 122, 2022.

\bibitem{liu2023false}
J.~Liu, R.~Zhang, S.~Szyller, K.~Ren, and N.~Asokan, ``False claims against
  model ownership resolution,'' in \emph{USENIX Security}, 2024.

\bibitem{liu2021secure}
X.~Liu, S.~Shao, Y.~Yang, K.~Wu, W.~Yang, and H.~Fang, ``Secure federated
  learning model verification: A client-side backdoor triggered watermarking
  scheme,'' in \emph{SMC}, 2021.

\bibitem{lukas2022sok}
N.~Lukas, E.~Jiang, X.~Li, and F.~Kerschbaum, ``Sok: How robust is image
  classification deep neural network watermarking?'' in \emph{S\&P}, 2022.

\bibitem{lv2023robustness}
P.~Lv, P.~Li, S.~Zhang, K.~Chen, R.~Liang, H.~Ma, Y.~Zhao, and Y.~Li, ``A
  robustness-assured white-box watermark in neural networks,'' \emph{IEEE
  Transactions on Dependable and Secure Computing}, 2023.

\bibitem{maini2020dataset}
P.~Maini, M.~Yaghini, and N.~Papernot, ``Dataset inference: Ownership
  resolution in machine learning,'' in \emph{ICLR}, 2020.

\bibitem{marcus1993building}
M.~P. Marcus, B.~Santorini, and M.~A. Marcinkiewicz, ``Building a large
  annotated corpus of {E}nglish: The {P}enn {T}reebank,'' \emph{Computational
  Linguistics}, vol.~19, no.~2, pp. 313--330, 1993.

\bibitem{merity2017pointer}
S.~Merity, C.~Xiong, J.~Bradbury, and R.~Socher, ``Pointer sentinel mixture
  models,'' in \emph{ICLR}, 2017.

\bibitem{minh2022explainable}
D.~Minh, H.~X. Wang, Y.~F. Li, and T.~N. Nguyen, ``Explainable artificial
  intelligence: a comprehensive review,'' \emph{Artificial Intelligence
  Review}, pp. 1--66, 2022.

\bibitem{naghiaei2022cpfair}
M.~Naghiaei, H.~A. Rahmani, and Y.~Deldjoo, ``Cpfair: Personalized consumer and
  producer fairness re-ranking for recommender systems,'' in \emph{SIGIR},
  2022.

\bibitem{noppel2023disguising}
M.~Noppel, L.~Peter, and C.~Wressnegger, ``Disguising attacks with
  explanation-aware backdoors,'' in \emph{S\&P}, 2023.

\bibitem{openai2023gpt}
OpenAI, ``Gpt-4 technical report,'' \emph{arXiv preprint arXiv:2303.08774},
  2023.

\bibitem{pan2022metav}
X.~Pan, Y.~Yan, M.~Zhang, and M.~Yang, ``Metav: A meta-verifier approach to
  task-agnostic model fingerprinting,'' in \emph{SIGKDD}, 2022.

\bibitem{paperno2016lambada}
D.~Paperno, G.~Kruszewski, A.~Lazaridou, Q.~N. Pham, R.~Bernardi, S.~Pezzelle,
  M.~Baroni, G.~Boleda, and R.~Fern{\'a}ndez, ``The lambada dataset: Word
  prediction requiring a broad discourse context,'' \emph{arXiv preprint
  arXiv:1606.06031}, 2016.

\bibitem{peng2022fingerprinting}
Z.~Peng, S.~Li, G.~Chen, C.~Zhang, H.~Zhu, and M.~Xue, ``Fingerprinting deep
  neural networks globally via universal adversarial perturbations,'' in
  \emph{CVPR}, 2022.

\bibitem{radford2019language}
A.~Radford, J.~Wu, R.~Child, D.~Luan, D.~Amodei, I.~Sutskever \emph{et~al.},
  ``Language models are unsupervised multitask learners,'' \emph{OpenAI blog},
  2019.

\bibitem{rana2015chi}
R.~Rana and R.~Singhal, ``Chi-square test and its application in hypothesis
  testing,'' \emph{Journal of Primary Care Specialties}, pp. 69--71, 2015.

\bibitem{ribeiro2016should}
M.~T. Ribeiro, S.~Singh, and C.~Guestrin, ``Why should i trust you? explaining
  the predictions of any classifier,'' in \emph{SIGKDD}, 2016.

\bibitem{sara2019image}
U.~Sara, M.~Akter, and M.~S. Uddin, ``Image quality assessment through fsim,
  ssim, mse and psnr—a comparative study,'' \emph{Journal of Computer and
  Communications}, vol.~7, no.~3, pp. 8--18, 2019.

\bibitem{selvaraju2017grad}
R.~R. Selvaraju, M.~Cogswell, A.~Das, R.~Vedantam, D.~Parikh, and D.~Batra,
  ``Grad-cam: Visual explanations from deep networks via gradient-based
  localization,'' in \emph{ICCV}, 2017.

\bibitem{shao2024fedtracker}
S.~Shao, W.~Yang, H.~Gu, Z.~Qin, L.~Fan, Q.~Yang, and K.~Ren, ``Fedtracker:
  Furnishing ownership verification and traceability for federated learning
  model,'' \emph{IEEE Transactions on Dependable and Secure Computing}, 2024.

\bibitem{sun2023deep}
Y.~Sun, T.~Liu, P.~Hu, Q.~Liao, S.~Ji, N.~Yu, D.~Guo, and L.~Liu, ``Deep
  intellectual property: A survey,'' \emph{arXiv preprint arXiv:2304.14613},
  2023.

\bibitem{tekgul2021waffle}
B.~G. Tekgul, Y.~Xia, S.~Marchal, and N.~Asokan, ``Waffle: Watermarking in
  federated learning,'' in \emph{SRDS}, 2021.

\bibitem{touvron2023llama}
H.~Touvron, T.~Lavril, G.~Izacard, X.~Martinet, M.-A. Lachaux, T.~Lacroix,
  B.~Rozi{\`e}re, N.~Goyal, E.~Hambro, F.~Azhar \emph{et~al.}, ``Llama: Open
  and efficient foundation language models,'' \emph{arXiv preprint
  arXiv:2302.13971}, 2023.

\bibitem{uchida2017embedding}
Y.~Uchida, Y.~Nagai, S.~Sakazawa, and S.~Satoh, ``Embedding watermarks into
  deep neural networks,'' in \emph{ICMR}, 2017.

\bibitem{waheed2024grove}
A.~Waheed, V.~Duddu, and N.~Asokan, ``Grove: Ownership verification of graph
  neural networks using embeddings,'' in \emph{S\&P}, 2024.

\bibitem{yan2023rethinking}
Y.~Yan, X.~Pan, M.~Zhang, and M.~Yang, ``Rethinking white-box watermarks on
  deep learning models under neural structural obfuscation,'' in \emph{USENIX
  Security}, 2023.

\bibitem{yang2023watermarking}
W.~Yang, S.~Shao, Y.~Yang, X.~Liu, X.~Liu, Z.~Xia, G.~Schaefer, and H.~Fang,
  ``Watermarking in secure federated learning: A verification framework based
  on client-side backdooring,'' \emph{ACM Transactions on Intelligent Systems
  and Technology}, 2023.

\bibitem{yao2023removalnet}
H.~Yao, Z.~Li, K.~Huang, J.~Lou, Z.~Qin, and K.~Ren, ``Removalnet: Dnn
  fingerprint removal attacks,'' \emph{IEEE Transactions on Dependable and
  Secure Computing}, 2023.

\bibitem{yao2024poisonprompt}
H.~Yao, J.~Lou, and Z.~Qin, ``Poisonprompt: Backdoor attack on prompt-based
  large language models,'' in \emph{ICASSP}, 2024.

\bibitem{yao2024prompt}
H.~Yao, J.~Lou, K.~Ren, and Z.~Qin, ``Promptcare: Prompt copyright protection
  by watermark injection and verification,'' in \emph{{S\&P}}, 2024.

\bibitem{yu2023self}
J.~Yu, H.~Yin, X.~Xia, T.~Chen, J.~Li, and Z.~Huang, ``Self-supervised learning
  for recommender systems: A survey,'' \emph{IEEE Transactions on Knowledge and
  Data Engineering}, 2023.

\bibitem{zhang2018protecting}
J.~Zhang, Z.~Gu, J.~Jang, H.~Wu, M.~P. Stoecklin, H.~Huang, and I.~Molloy,
  ``Protecting intellectual property of deep neural networks with
  watermarking,'' in \emph{AsiaCCS}, 2018.

\bibitem{zhu2015aligning}
Y.~Zhu, R.~Kiros, R.~Zemel, R.~Salakhutdinov, R.~Urtasun, A.~Torralba, and
  S.~Fidler, ``Aligning books and movies: Towards story-like visual
  explanations by watching movies and reading books,'' in \emph{ICCV}, 2015.

\end{thebibliography}
%



\appendix

\subsection{Additional Figures of the Experimental Results}
\label{apd:figs}

\begin{figure}[t]
    \centering
    \includegraphics[width=0.85\linewidth]{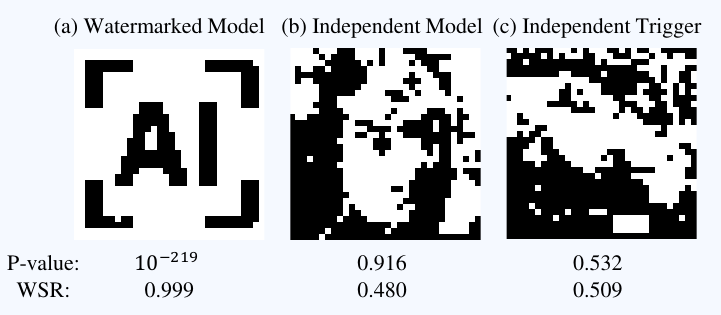}
    \caption{An example of the extracted watermarks with the watermarked model, independent model, and independent trigger.}
    \label{fig:independent}
    \vspace{-15pt}
\end{figure}

\begin{figure}[t]
    \centering
    \includegraphics[width=0.85\linewidth]{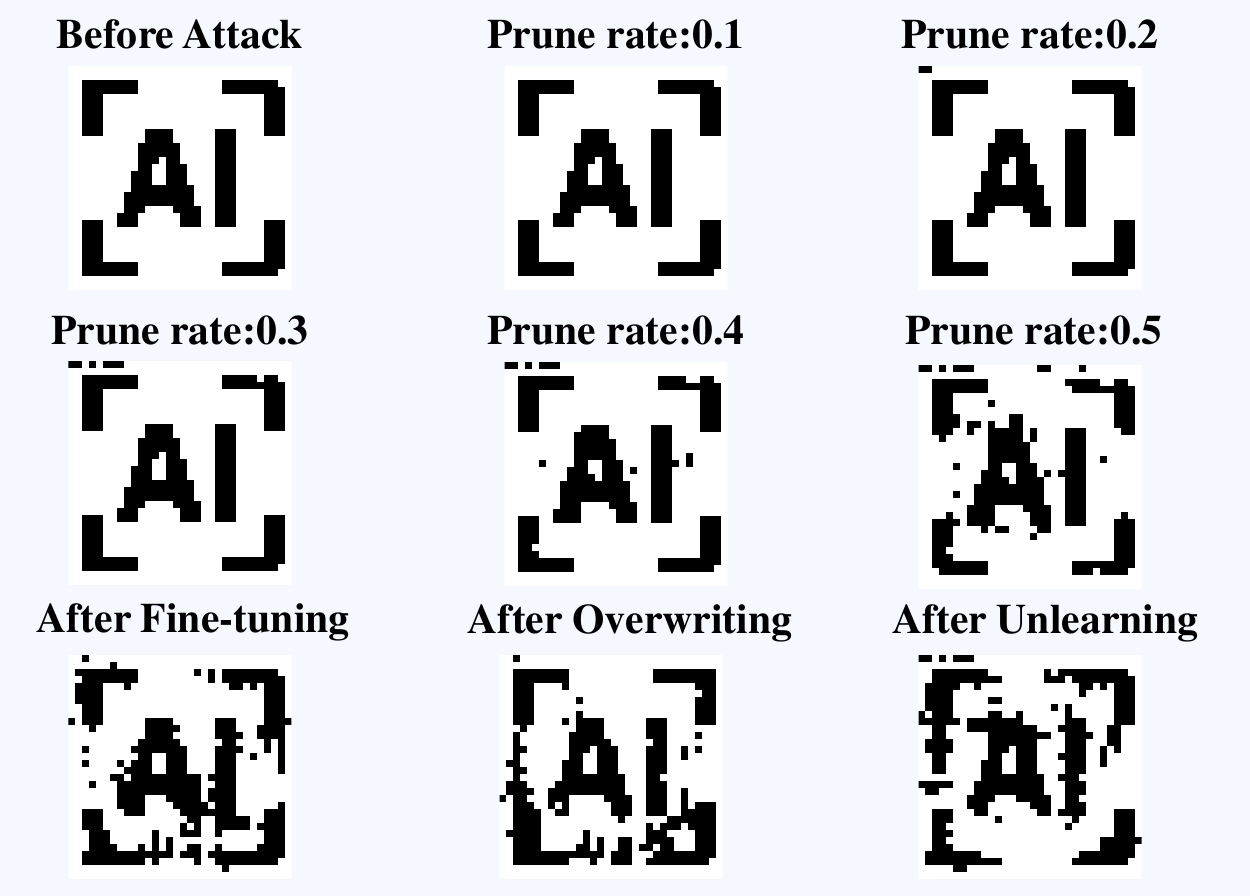}
    \caption{The visualization of the extracted watermarks before and after the watermark removal attacks.}
    \label{fig:attackvisual}
    \vspace{-20pt}
\end{figure}

Figure~\ref{fig:independent} shows an example of the extracted watermarks with the watermarked model, the independent model, and the independent trigger. Figure~\ref{fig:attackvisual} depicts the visualization of the extracted watermarks before and after the removal attacks.


\subsection{Implementation Details}
\label{apd:implementation}


\subsubsection{Implementation Details of the Experiments on Image Classification} 

We employ ResNet-18 trained on CIFAR-10 and a subset of ImageNet to embed the watermark. Given that the original architecture of ResNet-18 is primarily designed for ImageNet, we adjust the convolution kernel size of the first layer from $7\times 7$ to $3\times 3$ for training with CIFAR-10. The images in CIFAR-10 are $32 \times 32$, and the images in ImageNet are resized to $224\times 224$. The SGD optimizer is selected in our experiments with an initial learning rate of $5 \times 10^{-6}$. For hyper-parameter settings, the batch size is set to 128 for CIFAR-10 and 1024 for ImageNet. The value of $r_1$ in Eq.~(\ref{eq:wm}) is set to 1.0, while $\varepsilon$ in Eq.~(\ref{eq:hinge}) is set to $0.01$. To ensure determinism in the watermark extraction, we adopt a default setting where $k$ masks (with $k$ being the length of the watermark) are generated. In each mask, only one basic part is masked by setting its corresponding element as 0 and leaving all other elements as 1. 
The experiments are conducted utilizing four NVIDIA RTX 3090 GPUs.

\subsubsection{Implementation Details of the Experiments on Text Generation} 

\begin{figure*}[t]
    \centering
    \includegraphics[width=0.90\linewidth]{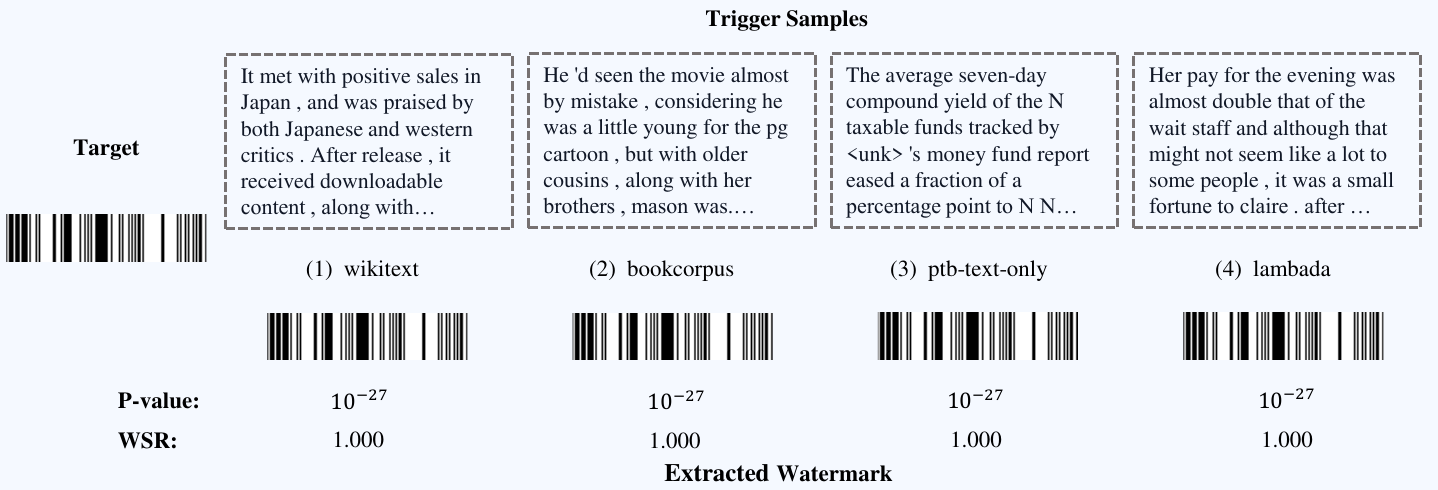}
    \caption{The trigger samples (on the upper row) used to watermark text generation models and the corresponding extracted watermark (on the bottom row). The target 1-D watermark (visualized as a bar code) is shown on the left.}
    \label{fig:text_trigger}
    \vspace{-20pt}
\end{figure*}

We fine-tune the GPT-2 model with four different datasets using the Adam optimizer. The learning rate is $3\times 10^{-4}$ and the batch size is $4$. Note that our goal is to evaluate the effectiveness of EaaW instead of training a high-performance model, considering the computational overhead, we set the max sequence length to be $128$ and we select $1,000$ sequences as the training set. We randomly select a sequence in the training set as the trigger sample. The examples are shown in Figure~\ref{fig:text_trigger}.
The experiments are carried out with two NVIDIA RTX A6000 GPUs.

\begin{table}[t]
    \tabcolsep=1.0mm
    \renewcommand{\arraystretch}{1.2}
    \centering
    \caption{Watermark success rate (WSR), the p-value, and test accuracy or perplexity (PPL) of applying EaaW to ResNet-101 and BERT.}
    \label{tab:moremodel}
    \scalebox{0.70}{
    \begin{tabular}{c|cccc|cccc}
    \hline
    \hline
        Model $\rightarrow$ & \multicolumn{4}{c|}{ResNet-101} & \multicolumn{4}{c}{BERT} \\
        Metric $\downarrow$ Length $\rightarrow$ & No WM & 64 & 256 & 1024 & No WM & 64 & 96 & 128\\
        \hline
        Test Acc. / PPL & 84.76 & 84.32 & 83.82 & 83.78 & 43.90 & 46.09 & 49.08 & 49.99\\
        p-value & / & $10^{-12}$ & $10^{-53}$ & $10^{-221}$ & / & $10^{-13}$ & $10^{-20}$ & $10^{-27}$\\
        WSR & / & 0.984 & 0.996 & 1.000 & / & 1.000 & 1.000 & 1.000\\
         \hline
         \hline
    \end{tabular}
    }
    \vspace{-15pt}
\end{table}

\begin{table}[t]
    \renewcommand{\arraystretch}{1.2}
    \centering
    \caption{Watermark success rate (WSR), the p-value, and test accuracy of ResNet-18 with different watermarks.}
    \label{tab:morewm}
    \scalebox{0.78}{
    \begin{tabular}{c|cccc}
    \hline
    \hline
        Dataset & Metric$\downarrow$ Watermark$\rightarrow$ & AI logo & Lock-like logo & Random\\
        \hline
        \multirow{3}{*}{CIFAR-10} & Test Acc. & 90.38 & 90.42 & 90.48\\
        & p-value & $10^{-222}$ & $10^{-220}$ & $10^{-222}$\\
        & WSR & 1.000 & 0.999 & 1.000\\
        \hline
        \multirow{3}{*}{ImageNet} & Test Acc. & 75.74 & 75.06 & 75.16\\
        & p-value & $10^{-222}$ & $10^{-220}$ & $10^{-222}$\\
        & WSR & 1.000 & 0.999 & 1.000\\
         \hline
         \hline
    \end{tabular}
    }
    \vspace{-15pt}
\end{table}

\begin{figure}[t]
    \centering
    \includegraphics[width=0.85\linewidth]{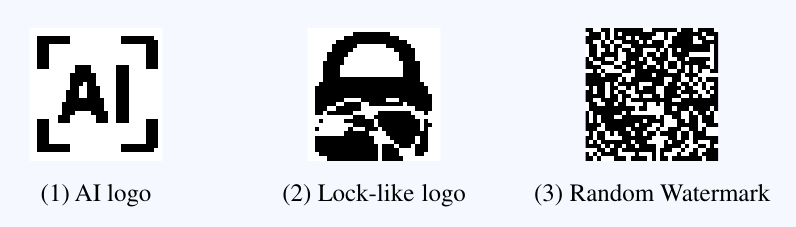}
    \caption{The visualization of the embedded watermarks.}
    \label{fig:morewm}
    \vspace{-20pt}
\end{figure}

\begin{table}[t]
    \tabcolsep=1mm
    \renewcommand{\arraystretch}{1.2}
    \centering
    \caption{The test accuracy (Test Acc.), the p-value, and the watermark success rate (WSR) using different watermark loss functions to embed the watermark into ResNet-18.}
    \label{tab:loss}
    \scalebox{0.74}{
    \begin{tabular}{c|ccccccc}
    \hline
    \hline
        Dataset & Length & Loss$\rightarrow$ & No WM  & Hinge-like & CE & MSE & SSIM\\
        \hline
        \multirow{9}{*}{ImageNet} & \multirow{3}{*}{64} & Test Acc. & 76.38 & \textbf{75.98} & 75.50 & 75.58 & 75.78\\
        & & p-value & / & $\bm{10^{-13}}$ & $\bm{10^{-13}}$ & $10^{-12}$ & $10^{-11}$\\
        & & WSR & / & \textbf{1.000} & \textbf{1.000} & 0.984 & 0.953\\
        \cline{2-8}
        & \multirow{3}{*}{256} & Test Acc. & 76.38 & \textbf{76.06} & 75.52 & 76.02 & 75.72\\
        & & p-value & / & $\bm{10^{-54}}$ & $\bm{10^{-54}}$ & $10^{-35}$ & $\bm{10^{-54}}$\\
        & & WSR & / & \textbf{1.000} & \textbf{1.000} & 0.922 & \textbf{1.000}\\
        \cline{2-8}
        & \multirow{3}{*}{1024} & Test Acc. & 76.38 & 75.74 & 75.16 & 75.74 & \textbf{76.08}\\
        & & p-value & / & $\bm{10^{-222}}$ & $10^{-188}$ & $10^{-91}$ & $10^{-19}$\\
        & & WSR & / & \textbf{1.000} & 0.970 & 0.860 & 0.725\\
         \hline
         \hline
    \end{tabular}
    }
    \vspace{-15pt}
\end{table}

\subsection{Additional Experiments}
\label{apd:ablation}

\subsubsection{Experiments with More Models}

In this section, we evaluate the effectiveness of EaaW with more models. We choose two models, ResNet-101~\cite{he2016deep} and BERT~\cite{devlin2018bert}, for discussion. ResNet-101 is a more powerful ResNet with 101 layers and BERT is another widely-used text generation model. We fine-tune the ResNet-101 and BERT with a subset of ImageNet and wikitext, respectively, and embed the watermarks via EaaW. The experimental results are shown in Table~\ref{tab:moremodel}. The p-values of both models are smaller than the significant level of $0.01$ and the WSRs are close to $1$. These results verify the effectiveness of EaaW on more current models.

\subsubsection{Experiments with Different Watermarks}

In this section, we evaluate EaaW by embedding different watermarks. We test two additional watermarks: the lock-like logo of NDSS and a random watermark (shown in Figure~\ref{fig:morewm}). As shown in Table~\ref{tab:morewm}, EaaW can successfully embed the watermark with nearly perfect WSRs. It validates the effectiveness of our EaaW regardless of targeted watermarks.

\subsubsection{Effect of Different Watermark Embedding Loss Functions}

In Section \ref{sec:embed}, we choose to use the hinge-like loss, which is also used in~\cite{shao2024fedtracker, fan2019rethinking}, as the watermark loss function $\mathcal{L}_2$. In this section, we conduct experiments to compare the effectiveness of utilizing other watermark loss functions for EaaW. The loss functions are listed below. 

\noindent\textbf{Cross Entropy Loss (CE)}: Since the elements in the watermark are either $1$ or $0$, the watermark embedding problem can be considered a binary classification problem~\cite{uchida2017embedding}. We can propose two different loss functions, cross-entropy loss and mean squared error loss. CE loss can be formalized as Eq.~(\ref{eq:ce}) where ${\tt sigmoid}(\cdot)$ is the sigmoid function. 
\begin{equation}
    \label{eq:ce}
    \mathcal{L}_2=-\sum_{i=1}^{k}\bm{\mathcal{W}}_i\log [{\tt sigmoid}(\bm{\mathcal{E}}_i)].
\end{equation}
\noindent\textbf{Mean Squared Error Loss (MSE)}: The mean squared error can also be used for binary classification problems. MSE loss can be formalized as Eq.~(\ref{eq:mse}). 
\begin{equation}
    \label{eq:mse}
    \mathcal{L}_2=\sum_{i=1}^k[\bm{\mathcal{W}}_i-{\tt sigmoid}(\bm{\mathcal{E}}_i)]^2.
\end{equation}
\noindent\textbf{Structure Similarity Index Measure Loss (SSIM)}: Structure similarity index measure can be used to measure the similarity of two images. The SSIM loss can be formalized as Eq.~(\ref{eq:ssim}). The detailed calculation of SSIM can be referred to~\cite{sara2019image}.
\begin{equation}
    \label{eq:ssim}
    \mathcal{L}_2=1- {\tt SSIM}[{\tt sigmoid}(\bm{\mathcal{E}}), \bm{\mathcal{W}}].
\end{equation}
We exploit the aforementioned watermark loss function together with hinge-like loss to embed the watermark into the ResNet-18 on ImageNet. The results in Table \ref{tab:loss} illustrate that in most cases, using hinge-like loss can achieve better effectiveness and harmlessness, while utilizing other watermark loss functions either cannot fully embed the watermark or cannot maintain the model's functionality.
When using CE and MSE, these two loss functions will make the absolute value of the explanation weights to infinity, causing the degradation of model functionality. Moreover, SSIM is relatively more complex than hinge-like loss so optimizing with SSIM loss is not easy in practice. Thus, using SSIM cannot acquire good effectiveness in embedding the watermark. In summary, we utilize the hinge-like loss as our watermark loss function. 

\begin{table}[t]
    \tabcolsep=1.5mm
    \renewcommand{\arraystretch}{1.2}
    \centering
    \caption{Watermark success rate (WSR) and test accuracy with different $\varepsilon$. We prune 40\% neurons to validate the resistance of the watermarks.}
    \label{tab:epsilon}
    \scalebox{0.79}{
    \begin{tabular}{c|ccccc}
    \hline
    \hline
        Model & Metric$\downarrow$ $\varepsilon$$\rightarrow$ & 0.1 & 0.01 (Ours) & 0.001 & 0.0001 \\
        \hline
        \multirow{3}{*}{ResNet-18} & Test Acc. & 75.48 & \textbf{75.72} & 75.44 & 75.32\\
        & WSR & \textbf{1.000} & \textbf{1.000} & 0.975 & 0.969\\
        & WSR after 40\% pruning & \textbf{0.995} & 0.980 & 0.773 & 0.620\\
         \hline
         \hline
    \end{tabular}
    }
    \vspace{-10pt}
\end{table}

\subsubsection{Effect of $\varepsilon$}

As shown in Eq.~(\ref{eq:hinge}), $\varepsilon$ is the hyper-parameter used in the watermark loss function, \ie, the hinge-like loss. $\varepsilon$ controls the resistance of the watermark against the removal attacks. We conduct the ablation study with $\varepsilon=0.1$ to $0.0001$ and apply a 40\%-pruning-attack to preliminarily validate the resistance of these embedded watermarks. ResNet-18 trained on ImageNet is used as the example model. The results in Table~\ref{tab:epsilon} demonstrate that a too small $\varepsilon$ may lead to poor resistance, while a too large $\varepsilon$ may compromise the utility of the models. To ensure both resistance and harmlessness, we choose to utilize $\varepsilon=0.01$ in our main experiments.




\begin{table}[t]
    \renewcommand{\arraystretch}{1.1}
    \centering
    \caption{The watermark success rate (WSR) using different numbers $c$ of masked samples during watermark embedding and watermark extraction.}
    \label{tab:label-only}
    \scalebox{0.76}{
    \begin{tabular}{c|cccccc}
    \hline
    \hline
        \multirow{2}{*}{Dataset} & \multirow{2}{*}{$c$ during embedding$\downarrow$} & \multicolumn{5}{c}{$c$ during extraction$\downarrow$} \\
        & & 256 & 512 & 1024 & 2048 & 4096\\
        \hline
        \multirow{5}{*}{ImageNet} & 256 & 0.566 & 0.590 & 0.605 & 0.594 & 0.633\\
        & 512 & 0.516 & 0.676 & 0.664 & 0.672 & 0.695\\
        & 1024 & 0.563 & 0.625 & 0.734 & 0.770 & 0.758\\
        & 2048 & 0.516 & 0.629 & 0.789 & 0.895 & 0.852\\
        & 4096 & 0.488 & 0.582 & 0.703 & 0.824 & 0.945\\
         \hline
         \hline
    \end{tabular}
    }
    \vspace{-15pt}
\end{table}

\subsection{Experiments in the Label-only Scenario}
\label{apd:label-only}

In this section, we investigate the effectiveness of EaaW in the label-only scenario, wherein the defender is restricted to obtaining only the predicted class rather than the logits. Consequently, for any masked samples, we assign a value of 1 to the corresponding element in the prediction vector $\bm{p}$ if it aligns with the correct class; otherwise, it is set to 0. While originally ranging between 0 and 1, these prediction logits $\bm{p}\in [0,1]$ are discretized as either 0 or 1 in this label-only scenario. As a result, there is a substantial reduction in available information for explaining data and models and extracting the watermark.

Therefore, in order to obtain an equivalent amount of information for watermark extraction, a straightforward approach is to increase the number of masked samples and queries for watermark extraction. Building upon this insight, we augment the quantity $c$ of masked samples during both watermark embedding and extraction processes. The experimental findings are presented in Table~\ref{tab:label-only}.

In this experiment, we aim to embed a $256$-bit watermark into the ResNet-18 model trained using the ImageNet dataset. However, when only a limited number of masked samples are utilized, EaaW fails to extract the watermark due to a WSR lower than $0.7$. Nevertheless, as the number of masked samples (c) surpasses $1024$, successful extraction of the watermark becomes feasible. These findings highlight that both watermark embedding and extraction in the label-only scenario necessitate an increased utilization of masked samples to ensure the effectiveness of EaaW. 

Furthermore, these results substantiate that augmenting the quantity of masked samples enables EaaW to function effectively even in scenarios where only labels are available. It also demonstrates the resistance of EaaW even under the worst-case attack: the adversary cannot remove the watermark without changing the predicted classes. 

\begin{table}[t]
    \tabcolsep=0.9mm
    \renewcommand{\arraystretch}{1.2}
    \centering
    \caption{The watermark success rate (WSR) and the accuracy when the adversary masks the input with different masking rates $\tau$ and different numbers $h$ of masks.}
    \label{tab:mask_input}
    \scalebox{0.68}{
    \begin{tabular}{c|cc|cc|cc|cc}
    \hline
    \hline
        $\tau\rightarrow$ & \multicolumn{2}{c|}{0.1\%} & \multicolumn{2}{c|}{1\%} & \multicolumn{2}{c|}{5\%} & \multicolumn{2}{c}{10\%}\\
        $h$$\downarrow$ Metric$\rightarrow$ & Test Acc. & WSR & Test Acc. & WSR & Test Acc. & WSR & Test Acc. & WSR\\
        \hline
        1 & 73.98 & 0.983 & 65.30 & 0.921 & 46.30 & 0.820 & 30.00 & 0.734\\
        3 & 74.12 & 0.987 & 65.68 & 0.963 & 46.94 & 0.822 & 29.90 & 0.714\\
        5 & 74.24 & 0.982 & 66.12 & 0.911 & 47.16 & 0.844 & 30.50 & 0.707\\
        10 & 74.32 & 0.990 & 66.00 & 0.971 & 47.38 & 0.825 & 30.34 & 0.742\\
         \hline
         \hline
    \end{tabular}
    }
    \vspace{-15pt}
\end{table}

\subsection{The Resistance to Adaptive Attacks via Modifying Inputs}
\label{apd:adaptive}

In Section~\ref{sec:adaptive}, we demonstrate the resistance of EaaW to adaptive attacks based on modifying the models. In this section, we further investigate the resistance to attacks that modify the inputs. In this type of attack, the adversary may add perturbations to the inputs to manipulate the explanations~\cite{ghorbani2019interpretation}. Specifically, given the input $\bm{x}$ and the model $f(x;\Theta)$, the adversary can randomly generate $h$ masks $M'=\{M'_i\}_{i=1}^h$ and utilize Eq.~(\ref{eq:mask}) to get the averaged output, as follows:
\begin{equation}
    \label{eq:mask}
    \bm{\bar{p}}=\frac{1}{h}\sum_{i=1}^hf(M'_i \otimes \bm{x}; \Theta).
\end{equation}

This adaptive attack is motivated by the fact that adding random masks may perturb the predictions and interfere with the watermark extraction since EaaW depends on the predictions of masked samples to extract the watermark. In particular, we define the proportion of $0$ in the masks $M'$ as the masking rate $\tau$ and implement the attacks using different $h$ and masking rates. As shown in Table~\ref{tab:mask_input}, the WSRs are still high even when setting a low masking rate $\tau$. In particular, as the masking rate $\tau$ increases, the utility of the model significantly drops but the WSRs are still higher than $0.70$, indicating the failure of this attack. Moreover, although using more masks (\ie, a large $h$) can slightly improve the test accuracy, it also raises the cost of inference. In summary, modifying the inputs will lead to a high inference overhead and a low utility. Accordingly, our EaaW resists this type of attack.

\begin{table}[t]
    \renewcommand{\arraystretch}{1.2}
    \centering
    \caption{The watermark success rate (WSR) using different numbers $c$ of masked samples during watermark embedding and watermark extraction.}
    \label{tab:maximum}
    \scalebox{0.78}{
    \begin{tabular}{c|ccc|ccc}
    \hline
    \hline
        Model$\rightarrow$ & \multicolumn{3}{c|}{ResNet-18} & \multicolumn{3}{c}{BERT}\\
        Metric$\downarrow$ Length$\rightarrow$ & 2025 & 3025 & 3600 & 150 & 170 & 185\\
        \hline
        accuracy/PPL & 74.38 & 73.36 & 74.62 & 50.45 & 56.66 & 57.81\\
        WSR & 0.997 & 0.962 & 0.605 & 1.000 & 1.000 & 1.000 \\
         \hline
         \hline
    \end{tabular}
    }
    \vspace{-10pt}
\end{table}

\subsection{Discussion on the Resistance to False Claim Attack}
\label{apd:falseclaim}

The false claim attack~\cite{liu2023false} is an improved version of the ambiguity attack~\cite{fan2019rethinking}. Both attacks aim to falsely claim to have ownership of another party's model. The difference is that the false claim attack attempts to find a transferable watermark certificate (\eg, trigger samples). Once the transferable certificate is registered, many third-party models trained afterward will be claimed as the intellectual properties of the adversary.

Liu \etal\cite{liu2023false} has demonstrated that existing backdoor-based model watermarking methods are vulnerable to the false claim attack since the adversary can easily construct some transferable adversarial examples. The vulnerability also stems from the zero-bit nature of the backdoor-based methods. On the contrary, since EaaW embeds a multi-bit watermark into the model, it is significantly more difficult for an adversary to conduct the false claim attack than zero-bit methods. We verify this statement both empirically (from Table~\ref{tab:cv_independent} and \ref{tab:nlp_independent}) and theoretically (in Section~\ref{sec:ambiguity}).

\subsection{Exploring the Maximum Embedded Watermark Length}

In this section, we investigate the maximum length of watermark that EaaW can embed. Our experiments are conducted using ResNet-18 and BERT models. The ResNet-18 model is trained on a subset of ImageNet and BERT is fine-tuned with the wikitext dataset. As depicted in Table~\ref{tab:maximum}, the maximum capacity of ResNet18 is greater than 3025 bits, but less than 3600. For BERT, we successfully embed a 185-bit watermark. Regrettably, further embedding of a larger-bit watermark is not feasible due to constraints in GPU memory.

\subsection{Analysis on the Efficiency of EaaW}
\label{apd:efficiency}

In this section, we analyze the efficiency of embedding the watermark using EaaW to illustrate that EaaW only has a slight increase in computational overhead compared with the backdoor-based methods. From Section \ref{sec:method}, the procedure of EaaW can be divided into several steps: \textbf{(1)} Prepare the $c$ masked data. \textbf{(2)} Input the $c$ masked data and get the prediction logits. \textbf{(3)} Using the metric function in Eq.~(\ref{eq:metric}) to calculate the metric vector $\bm{v}$. \textbf{(4)} Calculate the feature attribution matrix $\bm{W}$ through Eq.~(\ref{eq:normal}).

Compared to the backdoor-based method, we assume that the backdoor-based method needs to embed $c$ trigger samples into the model. First, the backdoor-based method also requires preparing $c$ trigger samples. The overhead of preparing the data can be neglected. Then, the backdoor-based method should optimize the model with these trigger samples. In one training iteration, the backdoor-based method involves one forward and one backward propagation. The overheads of these steps are close to Step 2\&3. From the above analysis, we can see that the only difference in the overhead between the EaaW and the backdoor-based method is Step 4. The equation of the Step 4 is as follows.
\begin{equation}
    \label{eq:normal-remark}
    \bm{W}=(M^TM+\lambda I)^{-1}M^T\bm{v},
\end{equation}
where $(M^TM+\lambda I)^{-1}M^T$ is constant. Therefore, in each iteration, the overhead of Step 4 is just one vector multiplication, which is negligible in the whole embedding process. In summary, the efficiency of EaaW is close to that of the backdoor-based model watermarking methods and our EaaW can efficiently embed the watermark into the models.

\subsection{Improving EaaW with Automated Hyperparameters Selection}

Since EaaW needs to preserve the utility of the model while embedding the watermark, the watermark embedding task can be defined as a multi-task optimization problem. In Eq.~(\ref{eq:wm}), we leverage a typical weighted sum optimization (WSO) that introduces a hyper-parameter $r_1$ to turn the multi-task optimization into the single-task optimization. Although our method is generally stable to the selection of $r_1$ as shown in Figure~\ref{fig:r1}, it can be further improved by automated hyperparameter selection techniques~\cite{krauss2024automatic}.

In this section, we implement the automated hyperparameter selection technique proposed in \cite{krauss2024automatic} and utilize the augmented Lagrangian method (ALM) to solve the watermark embedding problem. The results in Table~\ref{tab:auto} indicate that utilizing ALM can slightly improve the utility of the watermarked models. ALM is also free of hyperparameter selection. We will explore other optimization techniques in our future works. 

\begin{table}[t]
    \renewcommand{\arraystretch}{1.2}
    \centering
    \caption{The watermark success rate (WSR) and the test accuracy or perplexity (PPL) using our weighted sum optimization (Ours) or augmented Lagrangian method (ALM).}
    \label{tab:auto}
    \scalebox{0.82}{
    \begin{tabular}{c|cc|cc}
    \hline
    \hline
        Model$\rightarrow$ & \multicolumn{2}{c|}{ResNet-18} & \multicolumn{2}{c}{GPT-2}\\
        Metric$\downarrow$, Method$\rightarrow$ & Ours & ALM & Ours & ALM\\
        \hline
        accuracy/PPL & 75.52 & \textbf{75.64} & 48.99 & \textbf{47.94}\\
        WSR & \textbf{1.000} & 0.998 & \textbf{1.000}& \textbf{1.000} \\
         \hline
         \hline
    \end{tabular}
    }
    \vspace{-10pt}
\end{table}

\subsection{Potential Limitations}

Firstly, EaaW utilizes masked samples, which might cause misclassification and be further leveraged by the adversary as backdoor triggers. However, we argue that EaaW is still harmless. Specifically, \textbf{(1)} misclassifying masked samples is a pre-existing phenomenon as shown in Table~\ref{tab:mask_input}. EaaW doesn't introduce new threats. \textbf{(2)} The misclassification is untargeted which is less harmful. \textbf{(3)} When adding the masked samples to the training set, it can achieve an average of 99.04\% accuracy on the masked samples with 100\% WSR, indicating that EaaW can achieve high-level harmlessness.

Secondly, like other model watermarking methods, EaaW introduces extra overhead to embed the watermark. The time complexity is $O(c)$ where $c$ is the number of the masked data (\eg, 1024). However, as discussed in Appendix~\ref{apd:efficiency}, the overhead of EaaW is approximately equal to that of backdoor-based methods. Also, compared with the number of training samples which is usually larger than 50 thousand, the extra overhead is acceptable.

Thirdly, although EaaW has a negligible impact on the watermarked model, it is still an invasive model watermarking method. We will investigate how to design a non-invasive method, such as model fingerprinting, based on the insight of EaaW in our future work.

\end{document}